\documentclass[a4paper,11pt]{article}
\usepackage[a4paper,left=2.73cm,right=2.7cm,top=3cm,bottom=3.5cm]{geometry}

\usepackage{amsfonts,amsmath,amssymb,tabstackengine}
\usepackage[usenames, dvipsnames]{color}
\usepackage{array}

\usepackage{color, colortbl}
\definecolor{Gray}{gray}{0.95}

\usepackage[colorlinks=true,linktocpage=true,linkcolor=blue,citecolor=blue]{hyperref}
\usepackage{graphicx}

\addtocontents{toc}{\protect\setcounter{tocdepth}{2}}
\numberwithin{equation}{section}

\newcommand{\be}{\begin{equation}}
\newcommand{\ee}{\end{equation}}

\begin{document}

\begin{titlepage}

\thispagestyle{empty}

\begin{center}

{\LARGE \textbf{S-folds and (non-)supersymmetric Janus solutions}}

\vspace{40pt}
		
{\large \bf Adolfo Guarino}$\,^{a, b, c}$\,\,\,\,  \large{and}  \,\,\,\, {\large \bf Colin Sterckx}$\,^c$ 
		
\vspace{25pt}
		
$^a$\,{\normalsize  
Departamento de F\'isica, Universidad de Oviedo,\\
Avda. Federico Garc\'ia Lorca 18, 33007 Oviedo, Spain.}
\\[7mm]

$^b$\,{\normalsize  
Instituto Universitario de Ciencias y Tecnolog\'ias Espaciales de Asturias (ICTEA) \\
Calle de la Independencia 13, 33004 Oviedo, Spain.}
\\[7mm]

$^c$\,{\normalsize  
Universit\'e Libre de Bruxelles (ULB) and International Solvay Institutes,\\
Service  de Physique Th\'eorique et Math\'ematique, \\
Campus de la Plaine, CP 231, B-1050, Brussels, Belgium.}
\\[10mm]

\texttt{adolfo.guarino@uniovi.es} \,\, , \,\, \texttt{colin.sterckx@ulb.ac.be}

\vspace{20pt}

\vspace{20pt}
				
\abstract{
\noindent The S-fold description of Janus-type solutions of type IIB supergravity is investigated. This is done by first studying a $\,\textrm{U}(1) \times \textrm{U}(1)\,$ invariant sector of the four-dimensional dyonically-gauged $\,{[\,\textrm{SO}(1,1) \times \textrm{SO}(6)\,] \ltimes \mathbb{R}^{12}}\,$ maximal supergravity that arises upon reduction of type IIB supergravity on $\,\mathbb{R} \, \times \, \textrm{S}^{5}\,$. Two AdS$_{4}$ solutions preserving $\,\textrm{SU}(3)\,$ and $\,\textrm{SO}(6)\,$ gauge symmetry together with $\,\mathcal{N}=1\,$ and $\,\mathcal{N}=0\,$ supersymmetry are found within this sector. Fetching techniques from the E$_{7(7)}$ exceptional field theory, these solutions are uplifted to ten-dimensional S-folds of type IIB Janus-type solutions of the form $\,\textrm{AdS}_{4} \times \mathbb{R} \times \textrm{M}_{5}\,$. The solutions presented here are natural candidates for the holographic duals of three-dimensional $\,\mathcal{N}=1\,$ and $\,\mathcal{N}=0\,$ interface super-Yang--Mills theories with $\,\textrm{SU}(3)\,$ and $\,\textrm{SU}(4)\,$ internal symmetry.
}

\end{center}

\end{titlepage}

\tableofcontents

\hrulefill
\vspace{10pt}

\section{Motivation and summary}

Electromagnetic duality rotates the electric and magnetic fields of electromagnetism with an arbitrary angle $\,\omega\,$ \cite{Larmor:1928}. In the presence of charged matter, like electric charges or magnetic monopoles, this transformation is not a symmetry of Maxwell's equations and produces inequivalent physics. An analogue situation occurs in the context of maximal ($\mathcal{N}=8$) supergravity in four dimensions when a gauging of (a subgroup of) the $\,\textrm{E}_{7(7)}\,$ duality group is performed. At the bosonic level, the theory includes electric and magnetic non-Abelian gauge fields as well as charged scalar fields as matter. Electromagnetic duality again generates inequivalent physics \cite{deWit:2005ub}.

The prototypical example is the $\textrm{SO}(8)$-gauged supergravity that arises from the reduction of eleven-dimensional supergravity on a seven-sphere $\,\textrm{S}^7\,$ \cite{deWit:1982ig} and describes the near-horizon region of M2-branes. The action of electromagnetic duality leads to a one-parameter family of inequivalent theories depending on a continuous deformation parameter $\,\omega=\arg(1+ic)\,$ with $\,c \in [0,\sqrt{2}-1]\,$ \cite{Dall'Agata:2012bb}. The undeformed theory at $\,c=0\,$ possesses a maximally supersymmetric AdS$_{4}$ solution that preserves the full $\,\textrm{SO}(8)\,$ gauge group. This solution has an uplift to the Freund-Rubin solution of 11D supergravity \cite{Freund:1980xh} and is AdS$_{4}$/CFT$_{3}$ dual \cite{Maldacena:1997re} to the three-dimensional ABJM theory \cite{Aharony:2008ug} at low ($k = 1,2$) Chern--Simons levels $\,k\,$ and $\,-k\,$. In addition, other AdS$_{4}$ solutions of the undeformed theory with $\,\mathcal{N}=1\,\cite{Warner:1983vz,Comsa:2019rcz} \,,\,2\,\cite{Warner:1983vz}\,$ supersymmetry have been found which preserve less supersymmetry and partially break the $\,\textrm{SO}(8)\,$ gauge group via the Higgs mechanism. Some of these less supersymmetric solutions have been shown to be dual to various mass deformations of ABJM theory \cite{Bobev:2009ms}. However, when turning on $\,c\neq0\,$, new AdS$_{4}$ solutions preserving $\,\mathcal{N}=1\,\cite{Borghese:2012zs} \,,\,3\,\cite{Gallerati:2014xra}\,$ supersymmetry appear. Therefore the question arises: does the electromagnetic deformation parameter $\,c\,$ possess a higher-dimensional origin and a holographic interpretation? This question has been investigated during the last years without a satisfactory answer, thus singling out electromagnetic duality as a well-defined and controllable path to enter the (swamp)land of non-geometric supergravities in an \mbox{M-theory} context. 

\begin{table}[t!]
\begin{center}
\scalebox{0.94}{
\renewcommand{\arraystretch}{1.5}
\begin{tabular}{!{\vrule width 1.5pt}c!{\vrule width 0.5pt}c!{\vrule width 1pt}c!{\vrule width 1pt}c!{\vrule width 1.5pt}}
\noalign{\hrule height 1.5pt}
11D supergravity  & ?  & (massive) Type IIA & Type IIB \\
\noalign{\hrule height 1.5pt}
\multicolumn{2}{!{\vrule width 1.5pt}c!{\vrule width 1pt}}{SO(8)}  &  ISO(7) & $[\textrm{SO}(1,1) \times \textrm{SO}(6)] \ltimes \mathbb{R}^{12}$ \\
\noalign{\hrule height 1pt}
$c=0$   &   $c\neq 0$ & $c\neq0$ & $c\neq0$   \\
\noalign{\hrule height 1pt}
$\mathcal{N}=8$ / $\textrm{SO}(8)$ \,\,\cite{deWit:1982ig}  &   $\mathcal{N}=8$ / $\textrm{SO}(8)$ \,\,\cite{Dall'Agata:2012bb} &  &   \\[2mm]
   &    &  &  $\mathcal{N}=4$ / $\textrm{SO}(4)$  \,\,\cite{Gallerati:2014xra}   \\
   & $\mathcal{N}=3$ / $\textrm{SO}(4)$  \,\,\cite{Gallerati:2014xra} & $\mathcal{N}=3$ / $\textrm{SO}(4)$ \,\,\cite{Gallerati:2014xra} &    \\[2mm]
$\mathcal{N}=2$ / $\textrm{U}(3)$ \,\,\cite{Warner:1983vz}  &  $\mathcal{N}=2$ / $\textrm{U}(3)$  \,\,\cite{Borghese:2012zs} &  $\mathcal{N}=2$ / $\textrm{U}(3)$ \,\, \cite{Guarino:2015jca} &    \\[2mm]
$\mathcal{N}=1$ / $\textrm{G}_{2}$  \,\,\cite{Warner:1983vz}  &  $\mathcal{N}=1$ / $\textrm{G}_{2}$  \,\,\cite{Dall'Agata:2012bb} & $\mathcal{N}=1$ / $\textrm{G}_{2}$ \,\,\cite{Borghese:2012qm} &    \\[2mm]
\rowcolor{Gray}
   &  $\mathcal{N}=1$ / $\textrm{SU}(3)$ \,\,\cite{Borghese:2012zs} & $\mathcal{N}=1$ / $\textrm{SU}(3)$ \,\,\cite{Guarino:2015qaa} &  {\color{red}{$\mathcal{N}=1$ / $\textrm{SU}(3)$}} \,\,$[{\color{blue}{\textrm{here}}}]$ \\[0.5mm]
$\mathcal{N}=1$ / $\textrm{SU}(2)$ \,\, \cite{Comsa:2019rcz}   &    &  &   \\[2mm]
$\mathcal{N}=1$ / $\textrm{U}(1)^2$ \,\, \cite{Fischbacher:2009cj}   &    &  &   \\
\noalign{\hrule height 1.5pt}
\end{tabular}}
\caption{Supersymmetric AdS$_{4}$ solutions known up to date for the maximal supergravities with $\,\textrm{SO}(8)\,$, $\,\textrm{ISO}(7)\,$ and $\,[\textrm{SO}(1,1) \times \textrm{SO}(6)] \ltimes \mathbb{R}^{12}\,$ gauge groups. An entry of the form $\,\mathcal{N}\,/\,\textrm{G}_{0} \,\, [{\color{blue}{\textrm{ref}}}]\,$ in the table corresponds to an AdS$_{4}$ solution preserving $\,\mathcal{N}\,$ supersymmetries, $\,\textrm{G}_{0}\,$ residual gauge symmetry after Higgsing of the gauge group, and the reference where the solution was originally presented. It turns out that all the solutions with a given label $\,\mathcal{N}\,/\,\textrm{G}_{0}\,$ feature the same normalised mass spectra, which can be found in the corresponding reference, despite being solutions of different gauged supergravities and therefore having different uplifts to eleven-dimensional, (massive) type IIA or type IIB supergravity.}
\label{Table:vacua}
\end{center}
\end{table}

However more recent work has uncovered an unexpected connection with string theory. By studying the $\textrm{ISO}(7)$-gauged supergravity that arises from the reduction of type IIA ten-dimensional supergravity on a six-sphere $\,\textrm{S}^6\,$ and describes the near-horizon region of D2-branes, the electromagnetic deformation parameter $\,c\,$ was successfully identified with the Romans mass parameter of ten-dimensional type IIA supergravity and given a holographic interpretation as a Chern--Simons level in the dual CFT$_{3}$ \cite{Guarino:2015jca}. Unlike for the M-theory story, in this case the electromagnetic deformation turns out to be a discrete (on/off) deformation, \textit{i.e.}  $\,c=0 \,\,\textrm{or}\,\, 1\,$ \cite{Dall'Agata:2014ita}. Moreover there is no maximally supersymmetric AdS$_{4}$ solution in the massive IIA context. Still various examples of AdS$_{4}$ solutions with $\,\mathcal{N}=1\, \cite{Guarino:2015qaa}\,,\,2 \, \cite{Guarino:2015jca} \,,\,3\,\cite{Gallerati:2014xra}\,$ supersymmetry have been found in the $\textrm{ISO}(7)$-gauged supergravity when the electromagnetic deformation is activated. These findings have led to the discovery of a new gauge/gravity correspondence between AdS$_4$ backgrounds of massive IIA supergravity and super-Chern--Simons-matter theories with simple gauge group $\,\textrm{SU}(N)\,$ and level $\,k\,$ given by the Romans mass parameter \cite{Romans:1985tz}.

It is then natural to ask the same questions about the higher-dimensional origin and the dual field theory interpretation of electromagnetic duality now in the context of type IIB supergravity. This issue has been much less investigated in comparison with the M-theory and type IIA counterparts. Nonetheless the $[\textrm{SO}(1,1) \times \textrm{SO}(6)] \ltimes \mathbb{R}^{12}$-gauged supergravity has recenly been connected to a reduction of ten-dimensional type IIB supergravity on $\,\mathbb{R} \,\times\, \textrm{S}^{5}$~\cite{Inverso:2016eet}. In this case the electromagnetic deformation $\,c\,$ is again a discrete (on/off) parameter, and an AdS$_{4}$ solution preserving $\,\mathcal{N}=4\,$ supersymmetry and $\,\textrm{SO}(3) \times \textrm{SO}(3)\,$ gauge symmetry was found in \cite{Gallerati:2014xra}. This solution uplifts to S-fold backgrounds of type IIB supergravity of the form $\,\textrm{AdS}_{4} \times \Sigma \times \, \textrm{S}^2 \times \textrm{S}^2\,$ \cite{Inverso:2016eet}, which were shown to fall into the class of $\,\mathcal{N}=4\,$ Janus solutions presented in \cite{Assel:2011xz} (as certain limits of the type IIB solutions in \cite{DHoker:2007zhm,DHoker:2007hhe}). These are the gravity duals of $\,\mathcal{N}=4\,$ interface super-Yang--Mills theories \cite{DHoker:2006qeo,Gaiotto:2008sd}.

The set of supersymmetric AdS$_{4}$ solutions discussed above\footnote{An exhaustive classification of $\,\textrm{AdS}_{4}\,$ solutions in maximal gauged supergravity preserving $\,\mathcal{N}\ge 3\,$ was performed in \cite{Gallerati:2014xra}.} is summarised in Table~\ref{Table:vacua}, together with a new one (highlighted in red) to be presented in this work and uplifted along the lines of \cite{Inverso:2016eet} to S-fold backgrounds of type IIB supergravity. A quick inspection of \mbox{Table \ref{Table:vacua}} reveals that certain supersymmetric AdS$_{4}$ solutions, labelled by $\,\mathcal{N}\,/\,\textrm{G}_{0}\,$ in there, turn out to occur in different four-dimensional supergravities despite their rather distinct higher-dimensional origin\footnote{No higher-dimensional embedding is known for any of the AdS$_{4}$ solutions of the maximal $\textrm{SO}(8)$-gauged supergravity whenever $\,c \neq 0\,$.}. This motivates us to search for a simple solution occurring in the three different dyonically-gauged supergravities displayed in Table~\ref{Table:vacua}. In other words, a supersymmetric AdS$_{4}$ solution that enjoys an uplift to both type IIA/IIB supergravity and might also offer some new insight into a possible connection with eleven-dimensional supergravity.

To this end we will perform an exhaustive search for AdS$_{4}$ solutions preserving at least a $\,\textrm{U}(1)^2\,$ residual gauge symmetry in the $\,[\textrm{SO}(1,1) \times \textrm{SO}(6)] \ltimes \mathbb{R}^{12}\,$ dyonically-gauged maximal supergravity. This is the smallest residual symmetry preserved by a known supersymmetric solution of the undeformed SO(8)-gauged supergravity (see Table~\ref{Table:vacua}). Note also that, at the simple supersymmetric AdS$_{4}$ solution we are searching for, the $\,\textrm{U}(1)^2\,$ symmetry might be enhanced up to $\,\textrm{SU}(4)\sim\textrm{SO}(6)\,$ which is the largest compact subgroup that the $\,\textrm{SO}(8)\,$, $\,\textrm{ISO}(7)\,$ and $\,[\textrm{SO}(1,1) \times \textrm{SO}(6)] \ltimes \mathbb{R}^{12}\,$ gauge groups have in common. The result of our study is that such a simple supersymmetric AdS$_{4}$ solution exists with $\,\mathcal{N}=1\,$ supersymmetry and $\,\textrm{SU}(3)\,$ residual gauge symmetry (see Table~\ref{Table:vacua}). As a by-product we also discard the existence of an $\,\mathcal{N}=2$ / $\textrm{U}(3)\,$ solution in the $\,[\textrm{SO}(1,1) \times \textrm{SO}(6)] \ltimes \mathbb{R}^{12}$-gauged supergravity. Such a solution is nevertheless present in the $\,\textrm{SO}(8)\,$ and $\,\textrm{ISO}(7)\,$ theories and has been the subject of various holographic studies \cite{Benna:2008zy,Guarino:2015jca}. 

In the second part of the paper we uplift the new $\,\mathcal{N}=1\,/\,\textrm{SU}(3)\,$ AdS$_{4}$ solution of the $\,[\textrm{SO}(1,1) \times \textrm{SO}(6)] \ltimes \mathbb{R}^{12}$-gauged supergravity, as well as a companion $\,\mathcal{N}=0\,/\,\textrm{SO}(6)\,$ \textit{unstable} solution first reported in \cite{DallAgata:2011aa}, to non-compact Janus-type solutions of type IIB supergravity of the form $\,\textrm{AdS}_{4} \times \mathbb{R} \times \, \textrm{M}_{5}\,$. These solutions feature a varying axion-dilaton along a non-compact spatial direction we denote by $\,\eta\,$ (see \cite{Bak:2003jk,DHoker:2006vfr}). The internal space $\, \textrm{M}_{5}\,$ is either the round $\,\textrm{S}^5\,$ preserving $\,\textrm{SO}(6) \sim \textrm{SU}(4)\,$ or a squashed version of it, $\,\textrm{M}_5 = \mathbb{CP}_{2} \rtimes \textrm{S}^1\,$, preserving $\,\textrm{SU}(3) \subset \textrm{SU}(4)\,$. To do the uplift we fetch techniques from the $\,\textrm{E}_{7(7)}\,$ Exceptional Field Theory ($\textrm{E}_{7(7)}$-EFT) \cite{Hohm:2013uia} and perform a generalised Scherk--Schwarz reduction along the lines of \cite{Hohm:2014qga,Inverso:2016eet}. In this way we make the most of symmetries and geometry so that there is no need for guessing during the uplift process. The dependence of the type IIB fields on the coordinate $\,\eta\,$ turns out to be fully encoded into an $\,\textrm{SL}(2)_{\textrm{IIB}}\,$ twist matrix $\,A(\eta)\,$. This matrix is specified by the generalised Scherk--Schwarz ansatz, and can be used to compactify the $\,\eta\,$ direction and construct (compact) S-fold solutions for which the monodromy undergone by the axion-dilaton when going around $\,\eta\,$ is of hyperbolic type. These S-fold solutions are therefore natural candidates for the holographic duals of three-dimensional $\,\mathcal{N}=1\,$ and $\,\mathcal{N}=0\,$ interface super-Yang--Mills theories with $\,\textrm{SU}(3)\,$ and $\,\textrm{SU}(4)\,$ internal symmetry \cite{Clark:2004sb,DHoker:2006qeo}.

Our approach relies on, first, finding an AdS$_4$ solution with the appropriate (super) symmetries and, then, uplifting it directly to ten dimensions using the $\textrm{E}_{7(7)}$-EFT. This approach bypasses most of the difficulties that arise when one tries to, first, obtain a five-dimensional BPS domain-wall solution with AdS$_4$ slices in a five-dimensional supergravity and, then, uplift it to ten dimensions. This second approach makes analytic solutions difficult to obtain since, for example, in the case of supersymmetric domain-walls, one has to solve a complex system of first-order differential equations following from the vanishing of fermionic supersymmetry variations which usually can only be tackled numerically \cite{Clark:2005te,Suh:2011xc}. Our approach is purely algebraic once the information about the internal geometry is codified into the generalised Scherk--Schwarz reduction of $\textrm{E}_{7(7)}$-EFT.

The paper is organised as follows. In Section~\ref{Sec:4D} we construct the $\,\textrm{U}(1)^2\,$ invariant sector of the four-dimensional dyonically-gauged $\,{[\,\textrm{SO}(1,1) \times \textrm{SO}(6)\,] \ltimes \mathbb{R}^{12}}\,$ maximal supergravity and find two AdS$_{4}$ solutions. The first one is a new $\,\mathcal{N}=1\,$ supersymmetric solution preserving $\,\textrm{SU}(3)\,$ gauge symmetry. The second one is non-supersymmetric and preserves $\,\textrm{SO}(6)\,$ gauge symmetry. In Section~\ref{Sec:10D} we uplift these AdS$_{4}$ solutions to ten-dimensional S-folds of Janus-type solutions of type IIB supergravity with a hyperbolic monodromy. In Section~\ref{sec:conclusions} we present our conclusions. Two appendices accompany the main text. Appendix~\ref{App:Sasaki-Einstein} collects the geometrical data regarding the canonical Sasaki-Einstein structure on $\,\textrm{S}^5\,$ necessary to interpret the ten-dimensional type IIB solutions. Appendix~\ref{App:Type_IIB} summarises our conventions for type IIB supergravity and its equations of motion.

\section{New vacua of $\,{[\textrm{SO}(1,1) \times \textrm{SO}(6)] \ltimes \mathbb{R}^{12}}\,$ maximal supergravity}
\label{Sec:4D}

The dyonically-gauged maximal supergravity with $\,\textrm{G}={[\textrm{SO}(1,1) \times \textrm{SO}(6)] \ltimes \mathbb{R}^{12}} \subset \textrm{SL}(8)\,$ gauge group arises as the massless sector of the reduction of ten-dimensional chiral type IIB supergravity on $\,\mathbb{R} \times \textrm{S}^5\,$ \cite{Inverso:2016eet}. This theory corresponds to a gauging deformation of the ungauged maximal supergravity where the subgroup $\,\textrm{G}\subset \textrm{E}_{7(7)}\,$ of the global $\,\textrm{E}_{7(7)}\,$ duality group is promoted from global to local and therefore couples to the vector fields in the theory~\cite{deWit:2007mt}.

As shown in \cite{Inverso:2016eet,Dall'Agata:2014ita}, the group theoretical embedding
\begin{equation}
\textrm{SL}(8) \,\,\subset\,\, \textrm{E}_{7(7)} \,\,\subset\,\, \textrm{Sp}(56) \ ,
\end{equation}
where $\,\textrm{Sp}(56)\,$ is the group of electromagnetic transformations of the theory, allows for two inequivalent gaugings of $\,\textrm{G}\,$. These are specified by a discrete (on/off) parameter $\,c\,$ which specifies the choice of electromagnetic frame\footnote{Whenever non-zero, the parameter $\,c\,$ can be set to any value, \textit{i.e.} $\,c=1\,$, by a field redefinition \cite{Dall'Agata:2014ita}.}. As originally shown in \cite{Dall'Agata:2012bb}, the choice of electric/magnetic frame is of physical relevance once the gauging procedure is applied and, as a result, matter fields in the theory turn to be charged under the gauge group $\,\textrm{G}\,$.

\subsection{The $\,\mathcal{N}=8\,$ theory}

The dynamics of the maximal $\,\mathcal{N}=8\,$ theory is carried out by the field content of the supergravity multiplet. At the bosonic level it includes the metric field $\,g_{\mu\nu}\,$, $\,\textbf{28}\,$ electric $\,A_{\mu}{}^{AB}=A_{\mu}{}^{[AB]}\,$ and $\,\textbf{28'}\,$ magnetic $\,\tilde{A}_{\mu \, AB}=\tilde{A}_{\mu \, [AB]}\,$ vector fields, with $\,A=1,\dots,8\,$ being a fundamental index of $\,\textrm{SL}(8)\,$, and $\,70\,$ dynamical scalar fields serving as coordinates in the coset space $\,\textrm{E}_{7(7)}/\textrm{SU}(8)\,$. The vector fields\footnote{We use differential form notation for the various antisymmetric tensor fields in the theory.} can be assembled into an $\,\textrm{E}_{7(7)} \subset \textrm{Sp}(56) \,$ vector $\,A^{M}=(A^{AB} \,,\, \tilde{A}_{AB})\,$ with $\,M=1,\ldots,56\,$ being a fundamental $\,\textrm{E}_{7(7)}\,$ index. The scalar fields are encoded into a scalar-dependent matrix $\,M_{MN}\,$ which is in turn constructed as $\,M=\mathcal{V}\,\mathcal{V}^{t}\,$ using the coset element $\,\mathcal{V} \in \textrm{E}_{7(7)}/\textrm{SU}(8)\,$. This is the bosonic field content of maximal supergravity. However, in the presence of magnetic charges, an additional set of auxiliary two-form tensor fields not carrying independent dynamics must be introduced for the consistency of the Lagrangian formulation \cite{deWit:2005ub}.

The Lagrangian of the maximal four-dimensional supergravities was presented in \cite{deWit:2007mt}. It is specified in terms on an $\,\textrm{E}_{7(7)}$-valued embedding tensor $\,X_{MN}{}^{P} \in \textbf{912}\,$ subject to a set of quadratic constraints required by consistency of the gauging procedure (see also \cite{deWit:2005ub}). When the gauging is of the form $\,\textrm{G} \subset \textrm{SL}(8)\,$, as it is the case in the present work,  the theory is fully specified by two symmetric matrices $\,\eta_{AB}\,$ and $\,\tilde{\eta}^{AB}\,$. The former determines the piece of the gauge group that is spanned by electric vector $\,A_{\mu}{}^{AB}\,$ whereas the latter specifies the piece spanned by magnetic vectors $\,\tilde{A}_{\mu \, AB}\,$. This translates into a covariant derivative for the scalar fields of the form
\begin{equation}
\label{cov_deriv_N=8}
D M_{MN} = d M_{MN} \, - \, 2\, g \, A^{AB} \, X_{AB}{}_{(M}{}^{P} \, M_{N)P} - \, 2\, g \, \tilde{A}_{AB} \, X^{AB}{}_{(M}{}^{P} \, M_{N)P} \ ,
\end{equation}
with $\,g\,$ being the gauge coupling in the theory. Note that both electric and magnetic vectors appear in the gauge connection (\ref{cov_deriv_N=8}) thus specifying a so-called \textit{dyonic} gauging\footnote{Note that an electromagnetic frame in which the gauging is purely electric always exists by virtue of the quadratic constraints satisfied by the embedding tensor \cite{deWit:2005ub}.}. 
For the gauging $\,\textrm{G}={[\textrm{SO}(1,1) \times \textrm{SO}(6)] \ltimes \mathbb{R}^{12}}\,$ of interest in the present work one finds an embedding tensor of the form
\begin{equation}
\label{X-tensor}
\begin{array}{llllll}
{X_{[AB] [CD]}}^{[EF]} &=& - X_{[AB] \phantom{[EF]} [CD]}^{\phantom{[AB]} [EF]} &=& -8 \, \delta_{[A}^{[E} \, \eta_{B][C} \, \delta_{D]}^{F]}  &  , \\[2mm]
X^{[AB] \phantom{[CD]}[EF]}_{\phantom{[AB]}[CD]} &=& - {X^{[AB] [EF]}}_{[CD]} &=& -8 \, \delta_{[C}^{[A} \, \tilde{\eta}^{B][E} \, \delta_{D]}^{F]} &  ,
\end{array}
\end{equation}
with matrices given by\footnote{We have set $\,c=1\,$ in $\,\tilde{\eta}^{AB} = c \,\, \textrm{diag} (\, -1 \, , \, 0_{6}\, ,  \, 1 \,) \,$ without loss of generality (see footnote~$3$). Similarly we will also set $\,g=1\,$ which corresponds to an overall re-scaling of the embedding tensor.}
\begin{equation}
\eta_{AB} = \textrm{diag} (\, 0 \, , \, \mathbb{I}_{6}\, ,  \, 0 \,)
\hspace{8mm} \textrm{ and } \hspace{8mm}
\tilde{\eta}^{AB} = \textrm{diag} (\, -1 \, , \, 0_{6}\, ,  \, 1 \,) \ .
\end{equation}

Our goal in this section is to find new examples of anti-de Sitter (AdS$_{4}$) solutions in the four-dimensional supergravity. These correspond to maximally symmetric solutions in which vectors, as well as auxiliary tensor fields, must be set to zero. Before setting any field to zero, the Lagrangian is of the form
\begin{equation}
\label{L_N=8}
\mathcal{L}_{\mathcal{N}=8} = \left(\tfrac{R}{2} -V_{\mathcal{N}=8} \right) * 1 + \tfrac{1}{96} \,  \textrm{Tr}\left( D M \, \wedge * D M^{-1} \right) + \ldots \ ,
\end{equation}
where the ellipsis stand for Maxwell's and topological terms which do not play any role for AdS$_{4}$ solutions. The scalar potential in (\ref{L_N=8}) is determined in terms of the embedding tensor in (\ref{X-tensor}) and reads
\begin{equation}
\label{V_N=8}
V_{\mathcal{N}=8} =  \frac{g^{2}}{672} \,  {X_{MN}}^{R} \,  {X_{PQ}}^{S} \, M^{MP} \big( M^{NQ}  \, M_{RS} +   7 \,  \delta^{Q}_{R} \, \delta^{N}_{S}  \big) \ .
\end{equation}
We will search for new examples of AdS$_{4}$ solutions in the theory by direct minimisation of the scalar potential (\ref{V_N=8}). However, due to the large number of scalar fields in maximal supergravity, this problem becomes quite complex and some simplifications are required. To this end we will restrict to the subspace of the coset space $\,\textrm{E}_{7(7)}/\textrm{SU}(8)\,$ that is invariant under the action of a $\,\textrm{U}(1) \times \textrm{U}(1)\,$ subgroup with group-theoretical embedding
\begin{equation}
\label{U(1)xU(1)_embedding}
\textrm{U}(1) \times \textrm{U}(1)  \,\, \subset \,\,
\textrm{SU}(3)  \,\, \subset \,\,
\textrm{U}(1)_{\mathbb{U}} \times \textrm{SU}(3)  \,\, \subset \,\,
\textrm{SU}(4) \sim \textrm{SO}(6)  \,\, \subset \,\,
\textrm{G} \ .
\end{equation}
This subsector of the theory turns out to be an $\,\mathcal{N}=2\,$ supergravity coupled to various matter fields in the form of vector multiplets and a hypermultiplet, as we will explicitly show in the next section.

\subsection{$\textrm{U}(1)\times\textrm{U}(1)$ invariant sector}
\label{sec:N=2_model}

The $\,\textrm{U}(1) \times \textrm{U}(1)\,$ invariant sector of maximal supergravity with group-theoretical embedding
\begin{equation}
\label{U(1)xU(1)_non-compact_embedding}
\textrm{U}(1) \times \textrm{U}(1) \,\, \subset \,\,  \textrm{SO}(1,1) \times  \textrm{U}(1)_{\mathbb{U}} \times  \textrm{U}(1) \times \textrm{U}(1) \,\, \subset \,\, \textrm{SO}(1,1) \times  \textrm{SO}(6)  \ ,
\end{equation}
was investigated in \cite{Guarino:2017pkw,Guarino:2017jly} within the context of massive IIA and M-theory reductions on $\,\textrm{H}^{(p,q)}\,$ spaces. The scalar sector describes a coset geometry of the form
\begin{equation}
\label{scalar_geometry_N=2}
\mathcal{M}_{\textrm{scalar}} =  \mathcal{M}_{\textrm{SK}}  \times \mathcal{M}_{\textrm{QK}} \,\, \subset \,\, \frac{\textrm{E}_{7(7)}}{\textrm{SU}(8)} \ ,
\end{equation}
where
\begin{equation}
\label{scalar_geometry_factors_N=2}
\mathcal{M}_{\textrm{SK}} = \left[ \, \frac{\textrm{SL}(2)}{\textrm{SO}(2)} \, \right]^3
\hspace{8mm} \textrm{ and } \hspace{8mm}
\mathcal{M}_{\textrm{QK}} = \frac{\textrm{SU}(2,1)}{\textrm{SU}(2)\times\textrm{U}(1)}
\end{equation}
denote a special K\"ahler (SK) and a quaternionic K\"ahler (QK) geometry. The scalar geometry in (\ref{scalar_geometry_N=2}) involves a set of $\,t_{A}{}^{B}\,$ (with $\,t_{A}{}^{A}=0\,$) and $\,t_{ABCD}=t_{[ABCD]}\,$ generators of $\,\textrm{E}_{7(7)}\,$ in the SL(8) basis (see appendix in \cite{Guarino:2017jly} for conventions). The SK factor in (\ref{scalar_geometry_factors_N=2}) has associated generators given by
\begin{equation}
\label{Gener_SL2}
\begin{array}{ccll}
H_{\varphi_1}  & = &  {t_{4}}^{4} + {t_{5}}^{5} + {t_{6}}^{6} +  {t_{7}}^{7} - {t_{1}}^{1} - {t_{8}}^{8} - {t_{2}}^{2} - {t_{3}}^{3} & , \\[2mm]
H_{\varphi_2}  & = &  {t_{2}}^{2} + {t_{3}}^{3} + {t_{6}}^{6} +  {t_{7}}^{7} - {t_{1}}^{1} - {t_{8}}^{8} - {t_{4}}^{4} - {t_{5}}^{5} & , \\[2mm]
H_{\varphi_3}  & = &  {t_{2}}^{2} + {t_{3}}^{3} + {t_{4}}^{4} +  {t_{5}}^{5} - {t_{1}}^{1} - {t_{8}}^{8} - {t_{6}}^{6} - {t_{7}}^{7} & , \\[2mm]
g_{\chi_1} &=&  t_{1238} 
\hspace{3mm} , \hspace{3mm}
g_{\chi_2}  \,\,\,=\,\,\,  t_{1458}  
\hspace{3mm} , \hspace{3mm}
g_{\chi_3} \,\,\,=\,\,\,  t_{1678}  & ,
\end{array}
\end{equation}
whereas the QK factor in (\ref{scalar_geometry_factors_N=2}) is generated by
\begin{equation}
\label{Gener_SU3}
\begin{array}{ccll}

H_{\phi} &=& \tfrac{1}{2} \, ({t_{8}}^{8}-{t_{1}}^{1}) 
\hspace{3mm} , \hspace{2mm}
g_{\sigma} \,\,\, = \,\,\,  {t_{8}}^{1} & , \\[2mm]

g_{\zeta} & = &   t_{8357}   - t_{8346} - t_{8256} - t_{8247}  & ,  \\[2mm]

g_{\tilde{\zeta}}   & = &   t_{8246} - t_{8257} - t_{8347} - t_{8356}  & .
\end{array}
\end{equation}
As a result the coset representative of $\,\mathcal{M}_{\textrm{scalar}}\,$ in (\ref{scalar_geometry_N=2}) factorises as $\,{\mathcal{V}=\mathcal{V}_{\textrm{SK}} \times \mathcal{V}_{\textrm{QK}}}\,$ and is constructed upon the exponentiations
\begin{equation}
\label{coset_SL2xSU3}
\begin{array}{llll}
\mathcal{V}_{\textrm{SK}}  &=& \prod_{i} e^{-12 \, \chi_{i} \, g_{\chi_{i}}} \,\,\, e^{\frac{1}{4} \, \varphi_{i} \, H_{\varphi_{i}}} & ,  \\[2mm]
\mathcal{V}_{\textrm{QK}}  &=&  e^{\sigma \,  g_{\sigma}} \,\,\, e^{-6 \,(\zeta \, g_{\zeta} \,+\, \tilde{\zeta} \, g_{\tilde{\zeta}}) }  \,\,\, e^{-2 \, \phi \, H_{\phi}} & . 
\end{array}
\end{equation}
Finally the scalar matrix $\,M_{MN}\,$ in (\ref{L_N=8}) is obtained as $\,M=\mathcal{V} \, \mathcal{V}^{t}\,$ and, as anticipated, this sector of the theory can be recast in a canonical $\,\mathcal{N}=2\,$ form.

\subsubsection*{Canonical $\,\mathcal{N}=2\,$ formulation}

Omitting again Maxwell's and topological terms, which are incidentally irrelevant for the discussion of AdS$_{4}$ vacua, the Lagrangian of $\,\mathcal{N}=2\,$ supergravity coupled to $\,n_{v}\,$ vector multiplets and $\,n_{h}\,$ hypermultiplets takes the form
\begin{equation}
\label{Lagrangian_N2}
\mathcal{L}_{\mathcal{N}=2} =  \left( \frac{R}{2} - V_{\mathcal{N}=2} \right) *  1 - K_{i\bar{j}} \, Dz^{i} \wedge * D\bar{z}^{\bar{j}} - h_{uv} \, Dq^{u} \wedge * Dq^{v} + \ldots \ ,
\end{equation}
with $\,i=1,\ldots, n_{v}\,$ and $\,u=1,\ldots, 4\,n_{h}\,$. The matter multiplets of the $\,{\textrm{U}(1)\times\textrm{U}(1)}\,$ invariant sector of the maximal theory then correspond to three vector multiplets ($\,n_{v}=3\,$) and the universal hypermultiplet \cite{Cecotti:1988qn} ($\,n_{h}=1\,$) spanning the SK and QK geometries in (\ref{scalar_geometry_factors_N=2}), respectively. 

The scalar kinetic terms in (\ref{Lagrangian_N2}) can be read off from the scalar geometry data. The SK manifold in (\ref{scalar_geometry_factors_N=2}) is parameterised by three complex scalars $\,z^{i}=-\chi_{i} + i e^{-\varphi_{i}}\,$ serving as coordinates on the metric
\begin{equation}
\label{dSK}
ds_{\textrm{SK}}^2 = K_{i\bar{j}} \, dz^{i} d\bar{z}^{\bar{j}} = \frac{1}{4} \sum_{i}\frac{dz^{i} \, d\bar{z}^{\bar{i}}}{(\textrm{Im} z^{i})^2} 
\hspace{8mm} \textrm{ with } \hspace{8mm}
K_{i \bar{j}} = \partial_{z^{i}} \partial_{\bar{z}^{\bar{j}}} K \ ,
\end{equation}
where $\,K=-\log(i \left\langle X, \bar{X} \right\rangle)\,$ is the K\"ahler potential expressed in terms of the $\,\textrm{Sp}(2\,n_{v}+2)\,$ symplectic product
\begin{equation}
\left\langle X, \bar{X} \right\rangle=X^{M} \Omega_{MN} \bar{X}^{N}=X_{\Lambda}\bar{X}^{\Lambda}-X^{\Lambda}\bar{X}_{\Lambda}
\end{equation}
of holomorphic sections $\,X^{M}(z^{i})=(X^{\Lambda},F_{\Lambda})\,$ with $\,\Lambda=0,1,\ldots, n_{v}\,$. These sections are taken to satisfy $\,F_{\Lambda} = \partial \mathcal{F}/\partial X^{\Lambda}\,$ for a prepotential $\,\mathcal{F}\,$. We will take sections of the form
\begin{equation}
\label{Xsections_nv=3&nh=1}
(X^{0},X^{1},X^{2},X^{3},F_{0},F_{1},F_{2},F_{3}) =  (-z^{1} z^{2} z^{3}\,,\,-z^{1}\,,\,-z^{2}\,,\,-z^{3}\,,\,1\,,\, z^{2} z^{3}\,,\, z^{3} z^{1}\,,\, z^{1} z^{2}) \ ,
\end{equation}
which are consistent with the square-root prepotential
\begin{equation}
\label{F_prepot_nv=3&nh=1}
\mathcal{F} = -2 \sqrt{X^0 \, X^1 \, X^2 \, X^3} \ .
\end{equation}
For the hypermultiplet sector, the QK metric depends on coordinates $\,q^{u}=(\,\phi\,,\sigma\,,\,\zeta\,,\,\tilde{\zeta}\,)\,$ and takes the form
\begin{equation}
\label{dsQK_1}
ds_{\textrm{QK}}^2 = h_{uv} \, dq^{u} \, dq^{v} = d \phi \,  d \phi + \frac{1}{4} e^{4 \phi} \left[ d \sigma + \tfrac{1}{2}  \, \vec{\zeta} \,  \mathbb{C} \, d \vec{\zeta}  \right] \, \left[ d \sigma + \tfrac{1}{2}  \, \vec{\zeta} \,  \mathbb{C} \, d \vec{\zeta} \right] + \frac{1}{4} \, e^{2\phi} \,   d \vec{\zeta} \,  d\vec{\zeta}   \ ,
\end{equation}
with
\begin{equation}
\mathbb{C}=\left(  \begin{array}{cc} 0 & 1 \\ - 1 & 0\end{array} \right)
\hspace{10mm} \textrm{ and } \hspace{10mm}
\vec{\zeta}=(\zeta \,,\, \tilde{\zeta}) \ .
\end{equation}
We have verified that the scalar kinetic terms in (\ref{Lagrangian_N2}) exactly match the ones obtained from the maximal theory in (\ref{L_N=8}) when using the coset representatives in (\ref{coset_SL2xSU3}).

The dyonic gauging of the maximal theory gives rise, upon truncation to the $\,{\textrm{U}(1)\times\textrm{U}(1)}\,$ invariant sector, to a gauging of two Abelian hypermultiplet isometries with Killing vectors $\,k^{u}_{\alpha}=(k^{u}_{1}\,,\,k^{u}_{2})\,$. This gauging is specified by an $\,\mathcal{N}=2\,$ embedding tensor of the form
\begin{equation}
\label{Theta_tensor_N2}
\Theta_{M}{}^{\alpha} = \left(
\begin{array}{c}
\Theta_{\Lambda}{}^{\alpha} \\[2mm]
\hline\\[-2mm]
\Theta^{\Lambda \, \alpha} 
\end{array}\right)
\hspace{5mm} \textrm{ with } \hspace{5mm}
\Theta_{\Lambda}{}^{\alpha}
= \left(
\begin{array}{cc}
0 & 0 \\[2mm]
0 & 1\\[2mm]
0 & 1 \\[2mm]
0 & 1 
\end{array}\right) 
\hspace{5mm} \textrm{ and } \hspace{5mm}
\Theta^{\Lambda \, \alpha} 
= \left(
\begin{array}{cc}
1 & 0 \\[2mm]
0 & 0 \\[2mm]
0 & 0 \\[2mm]
0 & 0
\end{array}\right) \ ,
\end{equation}
where the index $\,M\,$ now runs over the vector fields of the $\,{\textrm{U}(1)\times\textrm{U}(1)}\,$ invariant sector:  electric $\,A^{\Lambda}\,$ and magnetic $\,\tilde{A}_{\Lambda}\,$ with $\,\Lambda=0,1,\ldots,n_{v}\,$. They are related with the vectors in the maximal theory as
\begin{equation}
\begin{array}{l}
A^{0}=A^{18}
\hspace{5mm} , \hspace{5mm} 
A^{1}=A^{23}
\hspace{5mm} , \hspace{5mm} 
A^{2}=A^{45}
\hspace{5mm} , \hspace{5mm} 
A^{3}=A^{67} \ , \\[2mm]
\tilde{A}_{0}=A_{18}
\hspace{5mm} , \hspace{5mm} 
\tilde{A}_{1}=A_{23}
\hspace{5mm} , \hspace{5mm} 
\tilde{A}_{2}=A_{45}
\hspace{5mm} , \hspace{5mm} 
\tilde{A}_{3}=A_{67} \ .
\end{array}
\end{equation}
The embedding tensor in (\ref{Theta_tensor_N2}) results in covariant derivatives of the form
\begin{equation}
\label{Dq}
\begin{array}{lllll}
D z^{i} &=& d z^{i} \ , \\[2mm]
D q^{u} &=& d q^{u} + g \, A^{M} \, \Theta_{M}{}^{\alpha} \, k_{\alpha}^{u} &=& d q^{u} + g \, \tilde{A}_{0} \,  \, k_{1}^{u} + g \, \mathcal{A}  \, k_{2}^{u} \ ,
\end{array}
\end{equation}
with $\,\mathcal{A} = \sum_{i} A^{i} \,$. There are three isometries of the QK geometry that play a prominent role in the reduction of type IIB supergravity on $\,\mathbb{R} \,\times\, \textrm{S}^5\,$. These are associated with Killing vectors that belong to the pair of duality-hidden symmetries \cite{Erbin:2014hsa} of the QK space
\begin{equation}
\label{Killing_vectors_sigma}
\begin{array}{lll}
k_{\sigma}&=&-\partial_{\sigma} \ , \\[2mm]
\widehat{k}_{\sigma}&=& \sigma \, \partial_{\phi} - (\sigma^2 - e^{-4 \phi} - U) \, \partial_{\sigma}  - \left[ \, \sigma \vec{\zeta} - \mathbb{C}  \, (\partial_{\vec{\zeta}} U)  \, \right]^{T} \,  \partial_{\vec{\zeta}} \ ,
\end{array}
\end{equation}
with
\begin{equation}
\label{Ufunc}
U = \frac{1}{16}  \,  |\vec{\zeta}|^4 \, + \, \frac{1}{2}   \, e^{-2 \phi } \, |\vec{\zeta}|^2  \  ,
\end{equation}
and the duality symmetry
\begin{equation}
\label{Killing_vectors_U}
\begin{array}{lll}
k_{\mathbb{U}}&=& \tilde{\zeta} \, \partial_{\zeta} - \zeta \, \partial_{\tilde{\zeta}}  \ .
\end{array}
\end{equation}
While $\,\widehat{k}_{\sigma} + k_{\sigma}\,$ and $\,k_{\mathbb{U}}\,$ -- the latter being rotations in the $\,(\zeta,\tilde{\zeta})$-plane -- are compact Abelian isometries of (\ref{dsQK_1}), the combination $\,\widehat{k}_{\sigma} - k_{\sigma}\,$ is non-compact. In the case of our type IIB model we find
\begin{equation}
\label{k1k2isometries}
k_{1} =\widehat{k}_{\sigma} -  k_{\sigma}   
\hspace{8mm} \textrm{ and } \hspace{8mm} 
k_{2}=k_{\mathbb{U}}  \ ,
\end{equation}
so that the $\,\mathcal{N}=2\,$ gauging in the $\,{\textrm{U}(1)\times\textrm{U}(1)}\,$ invariant sector is identified as
\begin{equation}
\label{gauging_N=2}
\textrm{G}=\textrm{SO}(1,1) \times \textrm{U}(1)_{\mathbb{U}} \ .
\end{equation}
This is the same gauge group that arises from M-theory models based on reductions on $\,\textrm{H}^{(7,1)}\,$ and $\,\textrm{H}^{(5,3)}\,$ hyperbolic spaces, although in those cases, only electric vectors happen to enter the gauging \cite{Guarino:2017jly}. For the type IIB model based on the reduction on $\,\mathbb{R} \times \textrm{S}^{5}\,$, the $\,\textrm{SO}(1,1)\,$ factor in (\ref{gauging_N=2}) is gauged by the magnetic graviphoton $\;\tilde{A}_{0}\,$ as it can be seen from (\ref{Dq}). Finally note also that $\,k_{1}\,$ does not vanish at any point in field space so that $\,\textrm{SO}(1,1)\,$ is always broken at any AdS$_{4}$ solution. On the contrary $\,\textrm{U}(1)_{\mathbb{U}}\,$ is preserved if $\,|\vec{\zeta}|=0\,$, namely, if $\,k_{2}=0\,$.

\subsection{Scalar potential and AdS$_4$ solutions}
\label{sec:AdS4_vacua}

In order to find AdS$_4$ solutions we will set to zero all the vector fields (and auxiliary tensors) in the $\,\mathcal{N}=2\,$ supergravity model and extremise the scalar potential $\,V_{\mathcal{N}=2}\,$ in (\ref{Lagrangian_N2}). To compute the latter, it turns convenient to introduce symplectic Killing vectors $\,\mathcal{K}_{M}\,$ and moment maps $\,\mathcal{P}_{M}^{x}\,$ defined as
\begin{equation}
\mathcal{K}_{M}  \equiv \Theta_{M}{}^{\alpha} \, k_{\alpha}
\hspace{10mm} \textrm{ and } \hspace{10mm}
\mathcal{P}_{M}^{x}  \equiv \Theta_{M}{}^{\alpha} \, P_{\alpha}^{x} \ ,
\end{equation}
so that symplectic covariance becomes manifest \cite{Klemm:2016wng}. The scalar potential is then expressed as \cite{Michelson:1996pn,deWit:2005ub}
\begin{equation}
\label{VN2}
\begin{array}{lll}
V_{\mathcal{N}=2} &=&  4 \,  \mathcal{V}^{M}  \, \bar{\mathcal{V}}^{N}    \, \mathcal{K}_{M}{}^{u}  \, h_{uv} \,  \mathcal{K}_{N}{}^{v} + \mathcal{P}^{x}_{M} \, \mathcal{P}^{x}_{N} \left( K^{i\bar{j}} \, D_{i}\mathcal{V}^{M} \, D_{\bar{j}} \bar{\mathcal{V}}^{N}  - 3 \, \mathcal{V}^{M} \, \bar{\mathcal{V}}^{N} \right) \ ,
\end{array}
\end{equation}
in terms of rescaled sections $\,\mathcal{V}^{M} \equiv e^{K/2} \, X^{M}\,$ and their K\"ahler derivatives $\,D_{i}\mathcal{V}^{M}=\partial_{z^{i}} \mathcal{V}^{M} + \frac{1}{2} (\partial_{z^{i}} K)\mathcal{V}^{M}\,$.  Therefore, in order to explicitly compute (\ref{VN2}), the only piece of information that is still left are the moment maps for the QK isometries in (\ref{Killing_vectors_sigma}) and (\ref{Killing_vectors_U}) entering the gauging (\ref{gauging_N=2}).

Following the general construction of \cite{Erbin:2014hsa}, the Killing vectors in (\ref{Killing_vectors_sigma}) are found to have associated moment maps of the form
\begin{equation}
P^{x}_{\sigma} =
\left(
\begin{array}{c}
0 \\[2mm]
0  \\[2mm]
-\frac{1}{2} \, e^{2 \phi } 
\end{array}
\right)
\hspace{6mm} \textrm{ and } \hspace{6mm}
\widehat{P}^{x}_{\sigma} =
\left(
\begin{array}{c}
- e^{-\phi } \, \tilde{\zeta } + e^{\phi }\,  \left( \, - \sigma \, \zeta  +\frac{1}{4}  \,  |\vec{\zeta}|^2 \,\tilde{\zeta } \, \right) \\[2mm]
e^{-\phi } \, \zeta   + e^{\phi }\,  \left( \, - \sigma \, \tilde{\zeta } - \frac{1}{4}   \,  |\vec{\zeta}|^2  \, \zeta \, \right) \\[2mm]
-\frac{1}{2} \, e^{-2 \phi } -\frac{1}{2} \, e^{2 \phi } \, \sigma ^2 -\frac{1}{32} \, e^{2 \phi } \,  |\vec{\zeta}|^4   + \frac{3}{4} \,  |\vec{\zeta}|^2 
\end{array}
\right) \ ,
\end{equation}
whereas the moment maps for the Killing vector in (\ref{Killing_vectors_U}) are given by the simpler expression
\begin{equation}
P^{x}_{\mathbb{U}} =
\left(
\begin{array}{c}
e^{\phi } \, \tilde{\zeta } \\[2mm]
- e^{\phi } \, \zeta \\[2mm]
1 - \frac{1}{4} \, e^{2 \phi } \, |\vec{\zeta}|^2
\end{array}
\right) \ .
\end{equation}
We have verified that the scalar potential in (\ref{VN2}) exactly matches the one in (\ref{V_N=8}) once the latter is restricted to the $\,\textrm{U}(1) \times \textrm{U}(1)\,$ invariant sector. Moreover, since the scalar potential is a gauge invariant quantity, it only depends on $\,\vec{\zeta}=(\zeta \,,\,\tilde{\zeta})\,$ through the combination $\,|\vec{\zeta}|\,$ which is invariant under the compact $\,\textrm{U}(1)_{\mathbb{U}}\,$. As a result, the angle $\,\gamma\,$ defined as $\,\vec{\zeta} = (\zeta \, , \, \tilde{\zeta}) = |\vec{\zeta}| \, (\cos \gamma \, , \, \sin \gamma)\,$ is identified with a Goldstone mode and corresponds with a flat direction in the scalar potential.

An explicit extremisation of the scalar potential in (\ref{VN2}) using \textsc{singular} \cite{DGPS} yields two inequivalent AdS$_4$ solutions\footnote{More concretely, we have decomposed the ideal $\langle \partial V_{\mathcal{N}=2} \rangle$ into primary factors using the Gianni-Trager-Zacharias (GTZ) primary decomposition algorithm, and then solved each of these (simpler) factors separately. We refer the reader to \cite{Gray:2006gn,Gray:2007yq} for a comprehensive review of Complex Algebraic Geometry techniques and their application to the study of string vacua. The specifics about the computer algebra system \textsc{singular} can be found at the official website \cite{DGPS}.}. The amount of supersymmetry preserved at a given AdS$_4$ solution can be determined by first evaluating the gravitino mass matrix \cite{Louis:2012ux}
\begin{equation}
\label{N=2_gravitino_mass}
S_{\mathcal{AB}}=\frac{1}{2} e^{K/2} X^{M}  \mathcal{P}^{x}_{M} (\sigma^{x})_{\mathcal{AB}} \ ,
\end{equation}
where $\,(\sigma^{x})_{\mathcal{AB}}\,$ are the Pauli matrices, and then checking how many (real) eigenvalues of $\,|\mathcal{S}|^2 = \mathcal{S}\,\mathcal{S}^{\dagger}\,$ take the same value as $\,\frac{1}{4} \, L^{-2}\,$. The number of such eigenvalues corresponds with the number of supersymmetries preserved at the AdS$_4$ solution.

\subsubsection*{$\mathcal{N}=1\,$ and $\,\textrm{SU}(3)$ invariant solution}

There is an $\,\mathcal{N}=1\,$ supersymmetric AdS$_4$ extremum of the potential for which the scalars in the vector multiplets are located at
\begin{equation}
\label{vacuum_N=1_SK}
\sum \textrm{Re}z^i = 0 
\hspace{8mm} , \hspace{8mm}
\textrm{Im}z^{i} = \frac{\sqrt{5}}{3} \ ,
\end{equation}
and those of the universal hypermultiplet take the values
\begin{equation}
\label{vacuum_N=1_QK}
e^{2 \phi} = \frac{6}{5} \, \frac{1}{\sqrt{1-\sigma^2} } 
\hspace{5mm} , \hspace{5mm}
\sigma \in (-1 , 1)
\hspace{5mm} , \hspace{5mm}
|\vec{\zeta}|^2 =  \tfrac{2}{3} \, \sqrt{1-\sigma^2} \ .
\end{equation}
Supersymmetry guarantees stability of this extremum as reflected by the normalised mass spectrum
\begin{equation}
\label{mass_N=1}
m^2  L^2 = 0 \,\, (\times 4) \,\, , \,\,  -2 \,\, (\times 2) \,\, , \,\,  4 \pm \sqrt{6} \,\, (\times 2) \ ,
\end{equation}
with $\,L^2=-3/V_{0}=\frac{25 \, \sqrt{5}}{54}\,$ being the AdS$_{4}$ radius. Additionally setting $\,\textrm{Re}z^{i} = 0\,$ in (\ref{vacuum_N=1_SK}) produces a residual symmetry enhancement to $\,\textrm{SU}(3)\,$ within the maximal theory despite the fact that the gauge group in (\ref{gauging_N=2}) is fully broken, in agreement with the group-theoretical embeddings in (\ref{U(1)xU(1)_embedding}) and (\ref{U(1)xU(1)_non-compact_embedding}). This new $\,\mathcal{N}=1\, / \,\textrm{SU}(3)\,$ AdS$_{4}$ solution has the same normalised scalar mass spectrum as its counterparts in the dyonically-gauged $\,\textrm{SO}(8)\,$ (see eq.~$(4.37)$ in \cite{Borghese:2012zs}) and $\,\textrm{ISO}(7)\,$ (see Table~$1$ in \cite{Guarino:2015qaa}) supergravities.

\subsubsection*{$\mathcal{N}=0\,$ and $\,\textrm{SO}(6)$ invariant solution}

There is a non-supersymmetric AdS$_4$ extremum of the potential for which the scalars in the vector multiplets are located at
\begin{equation}
\label{vacuum_N=0_SK}
\textrm{Re}z^{i} = \textrm{free} 
\hspace{8mm} , \hspace{8mm}
\textrm{Im}z^i = \frac{1}{\sqrt{2}} \ ,
\end{equation}
and those of the universal hypermultiplet take the values
\begin{equation}
\label{vacuum_N=0_QK}
e^{2 \phi} = \frac{1}{\sqrt{1-\sigma^2}} 
\hspace{5mm} , \hspace{5mm}
\sigma \in (-1 , 1)
\hspace{5mm} , \hspace{5mm}
|\vec{\zeta}|^2 = 0 \ .
\end{equation}
This extremum turns out to be perturbatively unstable as reflected by the normalised mass spectrum
\begin{equation}
\label{mass_N=0}
m^2  L^2 = 0 \,\, (\times 4) \,\, , \,\,  -3 \,\, (\times 2) \,\, , \,\,  6 \,\, (\times 2) \,\, , \,\,  -\frac{3}{4} \left( 1 - 2 \, \left(\sum \textrm{Re}z^i \right)^2 \right)  \,\, (\times 2) \ ,
\end{equation}
with $\,L^2=-3/V_{0}=\frac{3}{2 \, \sqrt{2}}\,$ being the AdS$_{4}$ radius. Note the presence of two normalised modes with $\,m^2 L^2 = -3 \,$ defining two particular directions in the scalar subspace spanned by $\,\textrm{Im}z^{i}\,$, and violating the Breitenlohner--Freedman (BF) bound $\,m^2 L^2 \ge -\frac{9}{4}\,$ for stability in AdS$_{4}$ \cite{Breitenlohner:1982bm}. Setting this time $\,\textrm{Re}z^{i} =0\,$ in (\ref{vacuum_N=0_SK}) makes the solution preserve a larger $\,\textrm{SO}(6)\sim \textrm{SU}(4)\,$ symmetry within the maximal theory, in agreement with the group-theoretical embeddings in (\ref{U(1)xU(1)_embedding}) and (\ref{U(1)xU(1)_non-compact_embedding}). This $\,\textrm{SO}(6)\,$ invariant solution, together with its normalised scalar mass spectrum, was reported in \cite{DallAgata:2011aa} (see solution {\large{x}} in Table~$4$ therein).

\section{Uplift to S-folds of type IIB supergravity}
\label{Sec:10D}

In this section we will construct the explicit uplift of the AdS$_{4}$ solutions in (\ref{vacuum_N=1_SK})-(\ref{vacuum_N=1_QK}) and (\ref{vacuum_N=0_SK})-(\ref{vacuum_N=0_QK}) to S-fold backgrounds of ten-dimensional type IIB supergravity. In particular, we will uplift the configurations with
\begin{equation}
\label{cond_Rez_i}
\textrm{Re}z^{i} = 0 \ ,
\end{equation}
for which we found an enhancement of symmetry from $\,\textrm{U}(1) \times \textrm{U}(1)\,$ to either $\,\textrm{SU}(3)\,$ or $\,\textrm{SO}(6)\,$. This symmetry enhancement allows us to work in the simpler $\,\textrm{SU}(3)\,$ invariant setup of \cite{Guarino:2015qaa}, which is related to the $\,\textrm{U}(1) \times \textrm{U}(1)\,$ invariant one upon the identification of the three vector multiplets
\begin{equation}
\label{SU(3)_invariance}
z^1=z^2=z^3\equiv z
\hspace{8mm} \textrm{ with } \hspace{8mm} 
z= -\chi + i \, e^{-\varphi} \ .
\end{equation}
Therefore the condition in (\ref{cond_Rez_i}) reduces to imposing $\,\chi=0\,$. It is worth emphasising that this is only consistent at the level of the four-dimensional equation of motion $\,\partial_{\chi}V_{\mathcal{N}=2}=0\,$ if the following relation holds
\begin{equation}
|\vec{\zeta}|^2 \,\, \left[  Y^2 - e^{4\phi} \, (1-\sigma^2) \right] = 0 \ ,
\end{equation}
with
\begin{equation}
\label{Y_def}
Y=1+\frac{1}{4} \, e^{2 \phi}  \, |\vec{\zeta}|^2 \ ,
\end{equation}
which is of course the case at both AdS$_{4}$ solutions. A direct evaluation shows that $\,Y=\frac{6}{5}\,$ for the $\,\mathcal{N}=1 \,/\, \textrm{SU}(3)\,$ solution with hypermultiplet scalars in (\ref{vacuum_N=1_QK}), and $\,Y=1\,$ for the $\,\mathcal{N}=0 \,/\, \textrm{SO}(6)\,$ solution with hypermultiplet scalars in (\ref{vacuum_N=0_QK}).

In order to perform the uplift we will fetch techniques from the $\,\textrm{E}_{7(7)}\,$ Exceptional Field Theory ($\textrm{E}_{7(7)}$-ExFT) constructed in \cite{Hohm:2013uia}. This is a field theory formally living $\,{(4+\textbf{56})}$-dimensional space-time which has a manifest $\,\textrm{E}_{7(7)}\,$ invariance under so-called  generalised diffeomorphisms acting on a $\textbf{56}$-dimensional internal space with coordinates $\,Y^{M}\,$. This generalised diffeomorphisms provide a unified description of both ordinary $\,\textrm{GL}(n)\,$ diffeomorphisms and internal gauge transformations for the various $p$-form fields of eleven-dimensional and type IIB supergravity. Importantly, the $\textrm{E}_{7(7)}$-ExFT requires a \textit{section constraint} for its consistent formulation which essentially reduces it to either eleven-dimensional supergravity in a $\,4+\textbf{7}\,$ dimensional split where internal $\,\textrm{GL}(7)\,$ diffeomorphisms are manifest, or type IIB supergravity in a $\,4+\textbf{6}\,$ dimensional split where internal $\,\textrm{GL}(6)\,$ diffeomorphisms are manifest. In this work we will be concerned with the type IIB solution of the section constraint and will perform the uplift of the previous AdS$_{4}$ solutions to type IIB supergravity by employing a so-called generalised Scherk--Schwarz (SS) ansatz for the various fields of $\textrm{E}_{7(7)}$-ExFT \cite{Hohm:2014qga}.

\subsection{Exceptional Field Theory and consistent truncations}

We are interested in uplifting AdS$_{4}$ solutions for which the four-dimensional scalars parameterising $\,M_{KL}(x) \in \textrm{E}_{7(7)}/\textrm{SU}(8)\,$ take constant vacuum expectation values and vectors, as well as tensor fields, vanish identically. The relevant fields in the $\textrm{E}_{7(7)}$-ExFT of \cite{Hohm:2013uia} are then the metric $\,g_{\mu\nu}(x,Y)\,$ and the generalised metric $\,\mathcal{M}_{MN} (x,Y)\,$. These fields are connected with the four-dimensional fields in (\ref{L_N=8}) via the generalised Scherk--Schwarz ansatz \cite{Hohm:2014qga}
\begin{equation}
\label{SS_ansatz}
\begin{array}{rll}
g_{\mu\nu}(x,Y) &=& \rho^{-2}(Y) g_{\mu\nu}(x) \ , \\[2mm]
\mathcal{M}_{MN} (x,Y) &=& U_M{}^K(Y) \, U_N{}^L(Y) \, M_{KL}(x) \ ,
\end{array}
\end{equation}
which is encoded into an $\,\textrm{SL}(8)\,$ twist matrix $\,U_{M}{}^{N}(Y)\,$ and an $\,\mathbb{R}^{+}\,$ scaling function $\,\rho(Y)\,$. In order for this ansatz to factorise out the dependence on the internal coordinates $\,Y^{M}\,$  at the level of the equations of motion and to give back the equations of motion of the four-dimensional theory, the twist matrix $\,U_{M}{}^{N}(Y)\,$ and the scaling function $\,\rho(Y)\,$ must fulfil the two conditions
\begin{equation}
\label{twist_conds}
\begin{array}{rcl}
\left. (U^{-1})_M{}^P \, (U^{-1})_N{}^Q \, \partial_P U_Q{}^K \right|_{\textbf{912}} &=& \tfrac{1}{7} \, \rho \, X_{MN}{}^{K} \ ,  \\[2mm]
\partial_N (U^{-1})_M{}^N - 3 \, \rho^{-1} \, \partial_N \rho \,  (U^{-1})_M{}^N &=& 2 \, \rho \, \vartheta_M \ ,
\end{array}
\end{equation}
where $\, X_{MN}{}^{K}\,$ is the embedding tensor in the four-dimensional gauged supergravity, $\,\vartheta_M\,$ is a constant (scaling) tensor, and $\,|_\textbf{912}\,$ is the projection onto the $\,\textbf{912} \in \textrm{E}_{7(7)}\,$ irreducible representation. For the twist matrix $\,U_{M}{}^{N}(Y)\,$ and the scaling function $\,\rho(Y)\,$ to be describing a background of type IIB supergravity, the dependence on the coordinates $\,Y^{M}\,$ of the generalised internal space must be such that the section constraint holds.

We will make use of various group-theoretical decompositions in order to establish a mapping between physical coordinates on the ordinary internal space and generalised coordinates. The first one involves the fundamental $\,\textrm{SL}(8)\,$ index and its decomposition $\,A=(1\,,\,m\,,\,8)\,$ with $\,m=2,\ldots 7\,$ being a $\,\textrm{GL}(6)\,$ fundamental index. Then the group-theoretical decomposition of the generalised coordinates $\,Y^{M}\,$ that is relevant for a type IIB interpretation of $\textrm{E}_{7(7)}$-ExFT is \cite{Hohm:2013uia}
\begin{equation}
\label{coordinates_GL(6)_branching}
\begin{array}{ccc}
\textrm{E}_{7(7)} & \supset & \textrm{GL}(6) \,\times \, \textrm{SL}(2)_{\textrm{IIB}}\,  \times \, \mathbb{R}^{+} \\[2mm]
\textbf{56} & \rightarrow & (\textbf{6},\textbf{1})_{+2} \, + \, (\textbf{6'},\textbf{2})_{+1}  \, + \, (\textbf{20},\textbf{1})_{0} \, + \, (\textbf{6},\textbf{2})_{-1} \, + \, (\textbf{6'},\textbf{1})_{-2} \\[2mm]
Y^{M} & \rightarrow & y^{m} \,\, + \,\, y_{m\alpha}  \,\, + \,\, y^{mnp}  \,\,  + \,\, y^{m\alpha} \,\, +  y_{m} \,\, 
\end{array}
\end{equation}
where $\,\textrm{GL}(6)\,$ accounts for ordinary six-dimensional internal diffeomorphisms and $\,\textrm{SL}(2)_{\textrm{IIB}}\,$ is the global symmetry of type IIB supergravity. In (\ref{coordinates_GL(6)_branching}) we have introduced a fundamental $\,\textrm{SL}(2)_{\textrm{IIB}}\,$  index $\,\alpha=1,8\,$ that is raised and lowered using $\,\epsilon_{18}=\epsilon^{18}=1\,$\footnote{We adopt conventions $\,v^{\alpha} = \epsilon^{\alpha \beta} \, v_{\beta}\,$ and $\,v_{\alpha} = v^{\beta}\,\epsilon_{\beta \alpha}\,$ for raising and lowering indices of $\,\textrm{SL}(2)_{\textrm{IIB}}\,$.}.  The fact that the six coordinates of the ordinary internal space of type IIB supergravity do not transform under $\,\textrm{SL}(2)_{\textrm{IIB}}\,$ suggests the choice of coordinates in $\,(\textbf{6},\textbf{1})_{+2} \,$, or equivalently in $\,(\textbf{6'},\textbf{1})_{-2}\,$, to be the physical coordinates. In this work we select
\begin{equation}
\label{phys_coords}
y^{m} \in (\textbf{6},\textbf{1})_{+2} \ .
\end{equation}
This choice solves the section constraint of $\,\textrm{E}_{7(7)}$-ExFT so that a type IIB generalised geometry in the sense of \cite{Coimbra:2011ky,Coimbra:2012af,Lee:2014mla} can be consistently defined on the tangent space of the ordinary internal space.

The background we are interested in is of the form $\,\mathbb{R} \times \textrm{S}^5 \,$ which further suggests a $\,\textrm{GL}(1) \,\times\, \textrm{GL}(5)\,$ splitting of coordinates $\,m=(i\,,\,7)\,$ with $\,i=2,\ldots6\,$ on the internal space. The specific mapping between $\,\textrm{GL}(6)\,$ coordinates in (\ref{coordinates_GL(6)_branching}), $\,\textrm{GL}(1) \,\times\, \textrm{GL}(5)\,$ coordinates and the generalised coordinates $\,Y^{M}=( Y^{AB} \,\, , \,\, Y_{AB})\,$ in the $\,\textrm{SL}(8)\,$ basis reads \cite{Inverso:2016eet}
\begin{equation}
\begin{array}{c|c|c|c|c}
 \,\,\,\,\,y^{m} \,\,\,\,& \,\,\,\,y_{m\alpha}  \,\,\,\, &\,\,\,\,\,\, y^{mnp} \,\,\,\, & \,\,\,\,\,\, y^{m\alpha} \,\,\,\, &  y_{m} \\[2mm]
\,\,y^{i} \,\,\,\,\,\,\,\,\,\,  y^{7} \,\,\,\,\,\,\,\, & \,\,y_{i\alpha} \,\,\,\,\,\,\,\,\,\,\,  y_{7\alpha} &  \,\,\,\,\,\,\,\,\,\,y^{ijk}  \,\,\,\,\,\,\,\,\,\,\,\,\,\,\,\,\,\,\, y^{ij7} \,\,\,\, & \,\,\,\, y^{i\alpha} \,\,\,\,\,\,\,\,\,   y^{7 \alpha} \,\,\,\, & y_{i} \,\,\,\,\,\,\,\,  y_{7} \\[2mm]
Y^{i7} \,\,\,\,\,\, Y_{18} \,\,\,\,\,\,\,\, & \,\,\,\,\,\,\, Y_{i \alpha} \,\,\,\,\, \epsilon_{\alpha \beta} \, Y^{\beta 7} &\,\,\,\,   \epsilon^{ijk j'k'}\, Y_{j'k'} \,\,\,\,\,\,\,  Y^{ij} \,\,\,\, &\,\,\,\,  Y^{i\alpha} \,\,\,\,\,  \epsilon^{\alpha \beta} \, Y_{\beta 7} \,\,\,\, & \,\,\,Y_{i7}  \,\,\,\,\,\,\,\, Y^{18} 
\end{array}
\end{equation}
rendering $\,y^{i}=Y^{i7}\,$ electric and $\,\tilde{y} \equiv y^{7}=Y_{18}\,$ magnetic in the generalised internal space. The twist matrix $\,U_{M}{}^{N}(y^{i},\tilde{y})\,$ and the scaling function $\,\rho(y^{i},\tilde{y})\,$ will determine the generalised SS reduction suitable to uplift the AdS$_{4}$ solutions previously found in a four-dimensional setup to backgrounds of type IIB supergravity. For the gauging $\,\textrm{G}=[ \textrm{SO}(1,1) \times \textrm{SO}(6) ] \ltimes \mathbb{R}^{12} \,$ the scaling function $\,\rho(y^{i},\tilde{y})\,$ is given by the product
\begin{equation}
\label{rho_func}
\rho(y^{i},\tilde{y}) = \mathring{\rho}(\tilde{y}) \, \hat{\rho}(y^{i}) \ ,
\end{equation}
where
\begin{equation}
\mathring{\rho}^{4} =1+ \, \tilde{y}^2 
\hspace{10mm} \textrm{ and } \hspace{10mm}
\hat{\rho}^{4} =1- \, |\vec{y}|^2 \ .
\end{equation}
Being valued in $\,\textrm{SL}(8) \in \textrm{E}_{7(7)}\,$, the (inverse) twist matrix $\,(U^{-1})_{M}{}^{N}(y^{i},\tilde{y})\,$ takes the block-diagonal structure
\begin{equation}
(U^{-1})_M{}^N \,=\, \begin{pmatrix}
(U^{-1})_{\left[AB\right]}{}^{\left[CD\right]} & 0 \\[2mm]
0 & (U^{-1})^{\left[AB\right]}{}_{\left[CD\right]} = U_{\left[CD\right]}{}^{\left[AB\right]} 
\end{pmatrix} \ ,
\end{equation}
with
\begin{equation}
(U^{-1})_{[AB]}{}^{[CD]} \,=\, (U^{-1})_A{}^C \, (U^{-1})_B{}^D - (U^{-1})_B{}^C \, (U^{-1})_A{}^D \ ,
\end{equation}
and
\begin{equation}
\label{U_inv_twist}
(U^{-1})_{A}{}^{B} \,=\, \left(\frac{\mathring{\rho}}{\hat{\rho}}\right)^\frac{1}{2} \, 
\left(
\begin{array}{cccc}
1 & 0 & 0 &  \mathring{\rho}^{-2}  \, \tilde{y} \\[2mm]
0 & \delta^{ij} + \hat{K} \, y^{i} \, y^{j} &  \hat{\rho}^2 y^{i} \,  & 0 \\[2mm]
0 &   \hat{\rho}^2 y^{j} \, \hat{K} \,  &  \hat{\rho}^4  & 0 \\[2mm]
\mathring{\rho}^{-2}  \, \tilde{y}  & 0 & 0 & \mathring{\rho}^{-4} (1+\tilde{y}^2) 
\end{array}
\right) \ .
\end{equation}
The function $\,\hat{K}(y^{i})\,$ entering the twist matrix in (\ref{U_inv_twist}) reduces in this case to a hypergeometric function \cite{Inverso:2016eet}
\begin{equation}
\hat{K} = -\phantom{}_2F_1\left(\, 1\,,\,2\,,\,\tfrac{1}{2} \, ; \, \,1-|\vec{y}|^2 \, \right) \ .
\end{equation}
The twist matrix $\,U_{M}{}^{N}\,$ and scaling function $\,\rho\,$ introduced above give rise, upon using (\ref{twist_conds}), to the embedding tensor $\,X_{MN}{}^{P}\,$ in (\ref{X-tensor}) and vanishing ``trombone" parameter $\,\vartheta_{M}\,$ \cite{LeDiffon:2008sh}.

Equipped with the $\,\textrm{SL}(8)\,$ twist matrix in (\ref{U_inv_twist}) and the scaling function in (\ref{rho_func}) it then becomes straightforward (but tedious) to uplift the AdS$_{4}$ solutions of Section~\ref{sec:AdS4_vacua} to solutions of $\,\textrm{E}_{7(7)}$-ExFT via the SS ansatz (\ref{SS_ansatz}). Once the explicit form of the external metric $\,g_{\mu\nu}(x^{\mu},y^{i},\tilde{y})\,$ and the generalised metric $\,\mathcal{M}_{MN}(x^{\mu},y^{i},\tilde{y})\,$ is determined, one can make use of the dictionary between the fields of $\,\textrm{E}_{7(7)}$-ExFT and the ones of type IIB supergravity \cite{Baguet:2015xha,Baguet:2015sma} to obtain an explicit type IIB background. To this end one must first perform a decomposition of the generalised metric $\,\mathcal{M}_{MN}\,$ under the relevant $\,\textrm{GL(6)} \times \textrm{SL}(2)_{\textrm{IIB}} \subset \textrm{E}_{7(7)}\,$ embedding in (\ref{coordinates_GL(6)_branching}) so that the dictionary reads
\begin{equation}
\label{uplift_formulas}
\begin{array}{rll}
G^{mn} &=& G^{\frac{1}{2}} \, \mathcal{M}^{mn} \ , \\[2mm]
\mathbb{B}_{mn}{}^\alpha &=& G^{\frac{1}{2}} \, G_{mp} \, \epsilon^{\alpha\beta} \, \mathcal{M}^{p}{}_{n\beta} \ , \\[2mm]
C_{klmn} - \frac{3}{2} \, \epsilon_{\alpha\beta} \, \mathbb{B}_{k\left[l\right.}{}^\alpha \,  \mathbb{B}_{\left.mn\right]}{}^\beta &=& - \frac{1}{2} \, G^{\frac{1}{2}} \,  G_{k\rho} \, \mathcal{M}^\rho{}_{lmn} \ ,\\[2mm]
m_{\alpha\beta} &=& \frac{1}{6} \, G \, \left( \, \mathcal{M}^{mn}\, \mathcal{M}_{m\alpha\,n\beta} + \mathcal{M}^m{}_{k\alpha} \, \mathcal{M}^k{}_{m\beta} \, \right) \ .
\end{array}
\end{equation}
The expressions in (\ref{uplift_formulas}) provide us with explicit formulas connecting the fields of $\,\textrm{E}_{7(7)}$-ExFT with the internal components of the fields of type IIB supergravity. These include the (inverse) metric $\,G^{mn}\,$, the $\,\textrm{SL}(2)_{\textrm{IIB}}\,$ doublet of two-form potentials $\,\mathbb{B}_{mn}{}^\alpha\,$, the four-form potential $\,C_{klmn}\,$ and the axion-dilaton matrix $\,m_{\alpha\beta}\,$ (see Appendix~\ref{App:Type_IIB}).

\subsection{Type IIB uplift of AdS$_{4}$ solutions}

In this section we perform the explicit uplift of the AdS$_{4}$ solutions of Section~\ref{sec:AdS4_vacua} to S-fold backgrounds of type IIB supergravity. As discussed at the beginning of the section, the AdS$_{4}$ solutions we are uplifting are compatible with the $\,\textrm{SU}(3)\,$ invariance in (\ref{SU(3)_invariance}) together with the further simplification of setting $\,\chi=0\,$. All the results we are presenting in this section are obtained in this simplified setup.

\subsubsection*{10D metric}

Before computing the ten-dimensional metric using (\ref{SS_ansatz}) and (\ref{uplift_formulas}) it will prove convenient to introduce some geometric data regarding the description of a five-sphere. Firstly the metric on the round $\,\textrm{S}^5\,$ and its inverse take the form
\begin{equation}
\hat{G}_{ij} = \delta_{ij} + \dfrac{\delta_{ik} \, \delta_{jl} \, y^{k} y^{l}}{1-y^{m} \, \delta_{mn} \, y^{n}}
\hspace{10mm} \textrm{ and } \hspace{10mm}
\hat{G}^{ij} = \delta^{ij} - y^{i} y^{j} \ .
\end{equation}
Secondly let us also introduce embedding coordinates $\,\mathcal{Y}_{\underline{m}}\,$ on $\,\mathbb{R}^{6}\,$ of the form
\begin{equation}
\mathcal{Y}_{\underline{m}}=\left\lbrace  y^{i} \, , \,  \mathcal{Y}_{7} \equiv \big(1- |\vec{y}|^2 \big)^{\frac{1}{2}} \right\rbrace 
\hspace{8mm} \textrm{ so that } \hspace{8mm}
\delta^{\underline{mn}} \, \mathcal{Y}_{\underline{m}} \, \mathcal{Y}_{\underline{n}} = 1 \ ,
\end{equation}
in terms of which the Killing vectors on $\,\textrm{S}^5\,$ are given by
\begin{equation}
\mathcal{K}_{\underline{mn}}{}^{i} \equiv \hat{G}^{ij} \, \partial_{j} \, \mathcal{Y}_{[\underline{m}} \, \mathcal{Y}_{\underline{n}]} 
=
\delta^{i}_{[\underline{m}} \, \mathcal{Y}_{\underline{n}]}  \ .
\end{equation}
Lastly we will also introduce the vector
\begin{equation}
\mathcal{K}^{i}  \equiv  \mathcal{K}_{\underline{mn}}{}^{i}  \,  J^{\underline{mn}} = J^{i \underline{m}}  \,  \mathcal{Y}_{\underline{m}} \ ,
\end{equation}
built from the real two-form $\,J\,$ of the $\textrm{SU}(3)$-holonomy of the K\"ahler cone $\,C(\textrm{S}^5)=\mathbb{R}^{+} \times \textrm{S}^5\,$ (see Appendix~\ref{App:Sasaki-Einstein}).

Using the uplifting formula for the internal metric in (\ref{uplift_formulas}) one finds an inverse six-dimensional metric of the form
\begin{equation}
\label{internal_metric}
\begin{array}{llll}
G^{11} & = &  \Delta \, \mathring{\rho}^4 \, M_{18\,18} \,\, = \,\, \Delta \, (1+\tilde{y}^{2}) \, e^{3\varphi} \ , \\[4mm]
G^{1k} & = &  \Delta \, \mathring{\rho}^2 \, \mathcal{K}_{\underline{ij}}{}^{k} \, M^{\underline{ij}}{}_{18} \,\, = \,\, 0  \ , \\[4mm]
G^{ij} & = & \Delta  \, \mathcal{K}_{\underline{kl}}{}^{i} \, \mathcal{K}_{\underline{mn}}{}^{j} \, M^{\underline{kl} \, \underline{mn} } \,\, = \,\,   \Delta \,  e^{\varphi}\, Y \,  \left[  \hat{G}^{ij} -  \left( 1- \frac{1}{Y}\right) \mathcal{K}^{i} \, \mathcal{K}^{j} \right]  \  , 
\end{array}
\end{equation}
where the blocks of the scalar matrix $\,M^{MN}\,$ that appear in (\ref{internal_metric}) are given by
\begin{equation}
\label{Mblocks_6Dmetric}
\begin{array}{llll}
M^{\underline{ij}\,\underline{kl}} &=& e^{\varphi} \,  \left[  (1-Y) \left( J^{\underline{ij}}J^{\underline{kl}} -3 \, J^{[\underline{ij}} J^{\underline{kl}]} \right) + 2 \, Y \delta^{\underline{k}[\underline{i}} \, \delta^{\underline{j}]\underline{l}} \right] \ , \\[2mm]
M^{\underline{ij}}{}_{18} &=& 0 \, \footnotemark \ , \\[2mm]
M_{18\,18} &=& e^{3 \varphi} \ ,
\end{array}
\end{equation}
\footnotetext{This block of $\,M^{MN}\,$ is proportional to the axion $\,\chi\,$ we are setting to zero by virtue of (\ref{cond_Rez_i}).}
and with 
\begin{equation}
\Delta = G^{\frac{1}{2}} \, \rho^2= \frac{e^{-\varphi}}{\sqrt{Y}} \ .
\end{equation}
We then find an internal six-dimensional metric of the form
\begin{equation}
\label{ds_6_0}
ds_{6}^2 \, =  \, \Delta^{-1}  \, e^{-3 \varphi} \, \dfrac{d\tilde{y}^{2} }{1+\tilde{y}^{2}} +  \Delta^{-1} \, e^{-\varphi} \, Y^{-1} \,  \left[   \, \hat{G}_{ij} + (Y-1) \, \mathcal{K}_{i} \, \mathcal{K}_{j} \right] dy^{i} \, dy^{j} \ ,
\end{equation}
where $\,\mathcal{K}_{i} \equiv \hat{G}_{ij} \, \mathcal{K}^{j}\,$. Performing a change of variable of the form 
\begin{equation}
\tilde{y}=\sinh \eta 
\hspace{8mm} \textrm{ with } \hspace{8mm}
\eta \in (-\infty \, , \, \infty) \ ,
\end{equation}
and using embedding coordinates $\,\mathcal{Y}^{\underline{m}}\,$, the metric in (\ref{ds_6_0}) takes the form 
\begin{equation}
\label{ds_6_1}
\begin{array}{lll}
ds_{6}^2 &=& \Delta^{-1} \, e^{-3 \varphi} \, d\eta^{2} +  \Delta^{-1} \, e^{-\varphi} \, Y^{-1}\,  \left[   \, \delta_{\underline{ij}} + (Y-1) \, J_{\underline{ki}} \, J_{\underline{lj}} \, \mathcal{Y}^{\underline{k}} \, \mathcal{Y}^{\underline{l}} \right] d\mathcal{Y}^{\underline{i}} \, d\mathcal{Y}^{\underline{j}} \ , \\[2mm]
 & = & \Delta^{-1} \, e^{-3 \varphi}   \, d\eta^{2} +  \Delta^{-1}  \, e^{-\varphi} \, Y^{-1}\,  \left[   \, ds^2_{\textrm{S}^5}+ (Y-1)  \, \boldsymbol{\eta}^2 \, \right] \ , 
 \end{array}
\end{equation}
so that
\begin{equation}
\label{ds_6}
\begin{array}{lll}
ds_{6}^2 &=& \Delta^{-1} \, e^{-3 \varphi}   \, d\eta^{2} +  \Delta^{-1}  \, e^{-\varphi} \, Y^{-1}\,  \left[   \, ds_{\mathbb{CP}^2}^2  + Y  \, \boldsymbol{\eta}^2 \, \right] \ , \\[2mm]
&=& \sqrt{Y} \, e^{-2 \varphi}  \, d\eta^{2} +   \frac{1}{\sqrt{Y}} \,  \left[   \, ds_{\mathbb{CP}^2}^2  + Y  \, \boldsymbol{\eta}^2 \, \right] \ .
 \end{array}
\end{equation}
In (\ref{ds_6_1}) and (\ref{ds_6}) we used the first relation in (\ref{SU2_structure}) and (\ref{metric_CP2xU1}) in order to exhibit the $\,\textrm{SU}(2)$-structure of the five-sphere $\,\textrm{S}^5\,$ when viewed as a Sasaki-Einstein manifold. 

Including also the four-dimensional (external) part of the geometry, we find a simple and non-singular $\,\textrm{AdS}_{4} \times \mathbb{R} \times \textrm{M}_{5}\,$ metric of the form
\begin{equation}
\label{metric_10D}
ds_{10}^2 =\tfrac{1}{2} \, \sqrt{Y} \, e^{\varphi}\, ds^2_{\textrm{AdS}_{4}} + \sqrt{Y} \, e^{-2 \varphi}  \, d\eta^{2} +   \frac{1}{\sqrt{Y}} \,  \left[   \, ds_{\mathbb{CP}^2}^2  + Y  \, \boldsymbol{\eta}^2 \, \right] \ ,
\end{equation}
in terms of the function $\,Y\,$ in (\ref{Y_def}) depending on the hypermultiplet scalars, and the scalar $\,\textrm{Im}z = e^{-\varphi}\,$ in the vector multiplet. Our choice of \textit{undeformed} frames for the metric (\ref{metric_10D}) reads
\begin{equation}
\begin{array}{llllll}
ds^2_{\textrm{AdS}_{4}} & : & \hat{e}^{0}=\dfrac{L}{r} \, dr & , & \hat{e}^{i}=\dfrac{L}{r}\, dx^{i} & \hspace{5mm} (i=1,2,3) \hspace{5mm} \textrm{ and } \hspace{5mm} \eta_{ij}=(-1,1,1)\\[2mm]
ds^2_{\mathbb{R}} & : & \hat{e}^{4} = d\eta  \\[2mm]
ds_{\mathbb{CP}^2}^2  & : & \hat{e}^{a}   \hspace{8mm}& & &  \hspace{5mm} (a=5,6,7,8) \\[2mm]
ds^2_{\textrm{S}^1} & : & \hat{e}^{9} = \boldsymbol{\eta}  \\[2mm]
\end{array}
\end{equation}
with $\,L\,$ being the AdS$_{4}$ radius at the four-dimensional solutions of section~\ref{sec:AdS4_vacua}, and where $\,\hat{e}^{a} \,$ and $\, \hat{e}^{9}\,$ describe a round $\,\textrm{S}^{5}\,$ as discussed in detail in Appendix~\ref{App:Sasaki-Einstein}. The volume form of the ten-dimensional space-time specified by the metric (\ref{metric_10D}) is then given by
\begin{equation}
\textrm{vol}_{10} = \tfrac{1}{4}  \, \Delta^{-1} \,\, \hat{e}^{0} \wedge \hat{e}^{1} \wedge \hat{e}^{2} \wedge \hat{e}^{3} \wedge \hat{e}^{4} \wedge \hat{e}^{5} \wedge \hat{e}^{6} \wedge \hat{e}^{7} \wedge \hat{e}^{8} \wedge \hat{e}^{9} \ .
\end{equation}

Two cases are of interest for the uplift of the AdS$_{4}$ solutions obtained in the previous section:
\begin{itemize}

\item[$i)\,$] For the $\,\mathcal{N}=1 \, / \, \textrm{SU}(3)\,$ solution in (\ref{vacuum_N=1_SK})-(\ref{vacuum_N=1_QK}) one has
\begin{equation}
Y=\frac{6}{5}
\hspace{8mm} \textrm{ and } \hspace{8mm}
e^{-\varphi} = \frac{\sqrt{5}}{3} \ .
\end{equation}
This makes the internal metric in (\ref{ds_6}) conform $\,\mathbb{CP}_{2} \rtimes \textrm{S}^1\,$ so that a $\,\textrm{U}(1)_{\beta}\,$ symmetry associated with $\,\boldsymbol{\eta}\,$ (see Appendix~\ref{App:Sasaki-Einstein}) is preserved together with the $\,\textrm{SU}(3)\,$ symmetry of $\,\mathbb{CP}_{2}\,$. We will see that this additional $\,\textrm{U}(1)_{\beta}\,$ symmetry is broken by the three-form fluxes, thus in agreement with the residual symmetry at the AdS$_{4}$ solution.\\[-2mm]

\item[$ii)\,$] For the $\,\mathcal{N}=0 \, / \, \textrm{SO}(6)\,$ solution in (\ref{vacuum_N=0_SK})-(\ref{vacuum_N=0_QK}) one has
\begin{equation}
Y = 1
\hspace{8mm} \textrm{ and } \hspace{8mm}
e^{-\varphi} = \frac{1}{\sqrt{2}} \ .
\end{equation}
In this case the round metric on $\,\textrm{S}^5\,$ is recovered with $\,\textrm{SO}(6)\,$ symmetry in agreement with the residual symmetry at the AdS$_{4}$ solution.

\end{itemize}

\subsubsection*{$B_{2}$ and $C_{2}$ potentials}

The $\,\textrm{SL}(2)_{\textrm{IIB}}\,$ doublet of two-form potentials $\,{\mathbb{B}^{\alpha}=(B_{2}\,,\,C_{2})}\,$ can be obtained from the second uplift formula in (\ref{uplift_formulas}). An explicit computation shows that
\begin{equation}
\begin{array}{lll}
\label{BC_uplift}
\mathbb{B}_{1j}{}^{\alpha} &=& 0 \ , \\[2mm]
\mathbb{B}_{ij}{}^{\alpha} &=& \Delta \, G_{ik} \, \mathcal{K}_{\underline{kl}}{}^{k} \, \partial_{j}\mathcal{Y}^{\underline{m}} \, \epsilon^{\alpha \beta} \, (A^{-1})^{\gamma}{}_{\beta} \, M^{\underline{kl}}{}_{\underline{m}\gamma} \ ,
\end{array}
\end{equation}
where the matrix
\begin{equation}
\label{Ainv_twist}
(A^{-1})^{\gamma}{}_{\beta} \equiv 
\left( 
\begin{array}{cc}
\sqrt{1+\tilde{y}^{2}} & -\tilde{y} \\[2mm]
-\tilde{y} & \sqrt{1+\tilde{y}^{2}}
\end{array}
\right) 
=
\left(
\begin{array}{cc}
\cosh\eta & -\sinh\eta \\[2mm]
- \sinh\eta & \cosh\eta
\end{array}
\right) \ ,
\end{equation}
is an $\,\textrm{SO}(1,1) \subset \textrm{SL}(2)_{\textrm{IIB}}\,$ element encoding the dependence of the two-form potentials $\,\mathbb{B}^{\alpha}\,$ on the direction $\,\eta\,$. The blocks of the scalar matrix $\,M^{MN}\,$ entering (\ref{BC_uplift}) are given by
\begin{equation}
\label{Mblocks_B2C2_potentials}
\begin{array}{llll}
M^{\underline{kl}}{}_{\underline{m}1} &=&  \frac{1}{2} \, e^{\varphi}  \,\, \left[ \,   j_{1} \, (\Omega^{I})^{\underline{kl}}{}_{\underline{m}} -   j_{2} \, (\Omega^{R})^{\underline{kl}}{}_{\underline{m}}  \, \right] \ , \\[2mm]
M^{\underline{kl}}{}_{\underline{m}8} &=&   \frac{1}{2} \, e^{2 \phi + \varphi}  \,\, \left[ \,   \tilde{\zeta} \, (\Omega^{R})^{\underline{kl}}{}_{\underline{m}} -   \zeta \, (\Omega^{I})^{\underline{kl}}{}_{\underline{m}}  \, \right]  \ ,
\end{array}
\end{equation}
in terms of the holomorphic three-form $\,\Omega = \Omega^{R} + i \,  \Omega^{I}\,$ of the $\textrm{SU}(3)$-holonomy of the K\"ahler cone $\,C(\textrm{S}^5)=\mathbb{R}^{+} \times \textrm{S}^5\,$ (see Appendix~\ref{App:Sasaki-Einstein}), and the scalar-dependent combinations 
\begin{equation}
j_{1}=\zeta \, Z + \tilde{\zeta} \, Y 
\hspace{5mm}  , \hspace{5mm}  
j_{2}=\tilde{\zeta} \, Z - \zeta \, Y
\hspace{8mm}  \textrm{ with } \hspace{8mm}  
Z=e^{2 \phi} \, \sigma \ .
\end{equation}
Inserting (\ref{Mblocks_B2C2_potentials}) into (\ref{BC_uplift}) one finds
\begin{equation}
\mathbb{B}_{ij}{}^{\alpha}  =  \tfrac{1}{2} \, e^{\varphi} \, \Delta  \, \epsilon^{\alpha \delta} \,  (A^{-1})^{\gamma}{}_{\delta} \, H_{\gamma \beta}\, G_{ik} \, (\Omega^{\beta})^{k}{}_{j}  = \tfrac{1}{2} \, Y^{-1}  \, \epsilon^{\alpha \delta} \,  (A^{-1})^{\gamma}{}_{\delta} \, H_{\gamma \beta} \, \hat{G}_{ik} \, (\Omega^{\beta})^{k}{}_{j}  \ ,
\end{equation}
in terms of the scalar-dependent matrix
\begin{equation}
\label{H_matrix}
H_{\gamma \beta} = 
\left(
\begin{array}{cc}
-j_{2} & j_{1} \\[2mm]
e^{2 \phi} \, \tilde{\zeta} & - e^{2 \phi} \, \zeta 
\end{array}
\right) \ ,
\end{equation}
and the geometric quantities
\begin{equation}
(\Omega^{\beta})^{i}{}_{j} \, \equiv \,  (\Omega^{R} , \Omega^{I})^{i}{}_{j} \, = \,  (\Omega^{R} , \Omega^{I})^{\underline{kl}}{}_{\underline{m}} \,\,  \mathcal{K}_{\underline{kl}}{}^{i} \,\, \partial_{j}\mathcal{Y}^{\underline{m}} \ ,
\end{equation}
which satisfy $\,\mathcal{K}_{k} \, (\Omega^{\beta})^{k}{}_{j} = 0\,$. Then, using the relation
\begin{equation}
\hat{G}_{ik} \, (\Omega^{\beta})^{k}{}_{j} \, dy^{i} \wedge  dy^{j}=  - (\Omega^{\beta})_{\underline{ijk}}\, \mathcal{Y}^{\underline{k}} \, d\mathcal{Y}^{\underline{i}} \wedge  d\mathcal{Y}^{\underline{j}} \ ,
\end{equation}
one arrives at the final expression
\begin{equation}
\mathbb{B}^{\alpha}  = \tfrac{1}{2} \, \mathbb{B}_{ij}{}^{\alpha} \, dy^{i} \wedge  dy^{j} =  - \tfrac{1}{4} \, Y^{-1}  \,  \epsilon^{\alpha\delta}  \, (A^{-t})_{\delta}{}^{\gamma} \, H_{\gamma \beta} \, (\Omega^{\beta})_{\underline{kij}}\, \mathcal{Y}^{\underline{k}} \, d\mathcal{Y}^{\underline{i}} \wedge  d\mathcal{Y}^{\underline{j}}  \ ,
\end{equation}
or, using the third relation in (\ref{SU2_structure}), the equivalent one
\begin{equation}
\label{(B,C)_potentials}
\mathbb{B}^{\alpha}  =  -  \tfrac{1}{2} \, Y^{-1}  \,  \epsilon^{\alpha\delta}  \, (A^{-t})_{\delta}{}^{\gamma} \, H_{\gamma \beta} \, \boldsymbol{\Omega}^{\beta}
\hspace{8mm} \textrm{ with } \hspace{8mm}
\boldsymbol{\Omega}^{\beta}  \equiv (\boldsymbol{\Omega}^{R} ,\boldsymbol{\Omega}^{I} ) \ .
\end{equation}

Taking an exterior derivative on (\ref{(B,C)_potentials}) and using the SU(2)-structure relations in (\ref{dSU(2)-structure}), one gets three-form fluxes $\,\mathbb{H}^{\alpha}=(H_{3},F_{3})\,$ of the form
\begin{equation}
\label{H_background}
\begin{array}{lll}
\mathbb{H}^{\alpha} &=& - \tfrac{3}{2} \, \,  Y^{-1}  \,  \epsilon^{\alpha\delta}  \, (A^{-t})_{\delta}{}^{\gamma} \, H_{\gamma \beta} \, (i \, \boldsymbol{\Omega} \wedge \boldsymbol{\eta} )^{\beta}  -  \tfrac{1}{2} \, Y^{-1}  \,  \epsilon^{\alpha\delta}  \, \partial_{\eta}(A^{-t})_{\delta}{}^{\gamma} \, H_{\gamma \beta} \, d\eta \wedge \boldsymbol{\Omega}^{\beta} \ , \\[2mm]
 & = & -  \tfrac{1}{2} \, Y^{-1}  \,  \epsilon^{\alpha\delta}  \, (A^{-t})_{\delta}{}^{\gamma} \left[ \, 3 \, H_{\gamma \beta} \, (i \, \boldsymbol{\Omega} \wedge \boldsymbol{\eta} )^{\beta} - \theta_{\gamma}{}^{\lambda} \,  H_{\lambda \beta} \, d\eta \wedge \boldsymbol{\Omega}^{\beta}\,  \right] \ ,
\end{array}
\end{equation}
in terms of the function $\,Y\,$ in (\ref{Y_def}) and $\,H_{\gamma \beta}\,$ in (\ref{H_matrix}) depending on the hypermultiplet scalars, and the matrix $\,A^{-1}(\eta)\,$ in (\ref{Ainv_twist}). In (\ref{H_background}) we have also introduced the constant matrix
\begin{equation}
\theta_{\gamma}{}^{\lambda} =
\begin{pmatrix}
0 & 1  \\[2mm]
1  & 0 
\end{pmatrix} \ .
\end{equation}

Two cases are again of interest for the uplift of the AdS$_{4}$ solutions obtained in the previous section:
\begin{itemize}

\item[$i)\,$] For the $\,\mathcal{N}=1 \, / \, \textrm{SU}(3)\,$ solution in (\ref{vacuum_N=1_SK})-(\ref{vacuum_N=1_QK}) one has
%
%
\begin{equation}
\label{Hmat_SU3}
H_{\alpha \beta} =\dfrac{Y \, |\vec{\zeta}| }{\sqrt{1-\sigma ^2}} 
\begin{pmatrix}
1 & \phantom{-}0  \\[2mm]
0  & -1 
\end{pmatrix}
\begin{pmatrix}
\sqrt{1-\sigma ^2}  \,\,  \cos\gamma  - \sigma \,\, \sin\gamma & \sqrt{1-\sigma ^2} \,\, \sin\gamma +  \sigma \,\, \cos\gamma  \\[2mm]
- \sin\gamma  & \cos\gamma 
\end{pmatrix}  \ ,
\end{equation}
with
\begin{equation}
\label{Y-zeta_N=1}
Y = \frac{6}{5}
\hspace{8mm} \textrm{ and }  \hspace{8mm}
|\vec{\zeta}|^2 =  \tfrac{2}{3} \, \sqrt{1-\sigma^2} \ .
\end{equation}
The $\,H_{\alpha \beta}\,$ matrix in (\ref{Hmat_SU3}) depends on $\,\sigma \in (-1 , 1)\,$ as well as on an angle $\,\gamma\,$ defined as $\,\vec{\zeta} = (\zeta \, , \, \tilde{\zeta}) = |\vec{\zeta}| \, (\cos \gamma \, , \, \sin \gamma)\,$ and parameterising a $\,\textrm{U}(1)_{\mathbb{U}}\,$ transformation. As a result there is a $\,(\sigma,\gamma)$-family of three-form fluxes (\ref{H_background}). The non-zero value $\,|\vec{\zeta}| \neq 0\,$ in (\ref{Y-zeta_N=1}) causes the Higgsing of the $\,\textrm{U}(1)_{\mathbb{U}}\,$ symmetry previously discussed in the four-dimensional context, where the flat direction in the scalar potential $\,\gamma\,$ was identified with a Goldstone mode. In addition, the three-form fluxes (\ref{H_background}) depend on the complex two-form $\,\boldsymbol{\Omega}\,$ that is charged (see Appendix~\ref{App:Sasaki-Einstein}) under the $\,\textrm{U}(1)_{\beta}\,$ preserved by the internal geometry (\ref{ds_6}). Therefore both $\,\textrm{U}(1)_{\mathbb{U}}\,$ and $\,\textrm{U}(1)_{\beta}\,$ symmetries are broken due to the background fluxes at the $\,\mathcal{N}=1\,$ solution.

\item[$ii)\,$] For the $\,\mathcal{N}=0 \, / \, \textrm{SO}(6)\,$ solution in (\ref{vacuum_N=0_SK})-(\ref{vacuum_N=0_QK}) one has $\,H_{\gamma \beta} = 0\,$ and therefore
\begin{equation}
\mathbb{H}^{\alpha}  =  0  \ ,
\end{equation}
as a consequence of the internal geometry being the round $\,\textrm{S}^5\,$.
 
\end{itemize}

\subsubsection*{$C_{4}$ potential}

An explicit computation using the third uplift formula in (\ref{uplift_formulas}) in combination with the first equation in (\ref{BC_uplift}) shows that the purely internal four-form potential is of the form
\begin{equation}
\begin{array}{rll}
C_{1jkl} &=& 0  \ , \\[2mm]
C_{ijkl} - \frac{3}{2} \, \epsilon_{\alpha \beta} \, \mathbb{B}_{i[j}{}^{\alpha} \, \mathbb{B}_{kl]}{}^{\beta} & = & - \frac{1}{2} \, \dfrac{\Delta}{\sqrt{1- |\vec{y}|^2}} \, \epsilon_{jkl k' l'} \, G_{i i'} \, \mathcal{K}_{\underline{mn}}{}^{i'} \,\, \mathcal{K}_{\underline{pq}}^{k' l'}  \,\, M^{\underline{mn} \,  \underline{pq}} \ ,
\end{array}
\end{equation}
where we have introduced the geometric tensor
\begin{equation}
\mathcal{K}_{\underline{pq}}^{k' l'} \,\, = \,\, \delta_{\underline{pq}}^{k' l'} + 2 \, \hat{K} \, \delta_{\underline{p}}^{k'} \, \mathcal{Y}_{\underline{q}} \, y^{l'} \ . 
\end{equation}
Substituting the first scalar-dependent block in (\ref{Mblocks_6Dmetric}) we arrive at
\begin{equation}
\label{c4_potential_0}
\begin{array}{lll}
C_{ijkl} - \tfrac{3}{2} \, \epsilon_{\alpha \beta} \, \mathbb{B}_{i[j}{}^{\alpha} \, \mathbb{B}_{kl]}{}^{\beta}  &=& \epsilon_{j'jklk'} \, \dfrac{1}{\sqrt{1- |\vec{y}|^2}} \, \dfrac{Y-1}{Y} \,  \, \hat{G}_{i\,i'}  \, \left[ J^{i'j'} \, J^{k'\underline{l}} \, \mathcal{Y}_{\underline{l}} +  \delta^{j'l'} \, \mathcal{K}^{i'} \, \mathcal{K}_{l'}\, y^{k'} \right] \\[4mm]
&+& \hat{C}_{ijkl}\ ,
\end{array}
\end{equation}
with
\begin{equation}
\label{c4_potential_round}
\begin{array}{lll}
\hat{C}_{ijkl} = \epsilon_{ijklk'} \dfrac{y^{k'}}{\sqrt{1- |\vec{y}|^2}}  \,  (1 + \hat{K}) \ .
\end{array}
\end{equation}
A careful analysis of the expression in (\ref{c4_potential_0}) reveals that the contribution
\begin{equation}
-\tfrac{3}{2} \, \epsilon_{\alpha \beta} \, \mathbb{B}_{i[j}{}^{\alpha} \, \mathbb{B}_{kl]}{}^{\beta} = -6 \, \frac{1-Y}{Y} \,  \epsilon_{\alpha \beta} \, \boldsymbol{\Omega}_{i[j}{}^{\alpha} \, \boldsymbol{\Omega}_{kl]}{}^{\beta} \ ,
\end{equation}
in the left hand side precisely cancels against the contribution coming from the first term in the right hand side so that
\begin{equation}
\label{c4_potential}
C_{ijkl} = \hat{C}_{ijkl} \ .
\end{equation}
The purely internal five-form flux then takes the form
\begin{equation}
dC = \boldsymbol{\Omega} \wedge \bar{\boldsymbol{\Omega}} \wedge \boldsymbol{\eta} = 4 \, Y^{\frac{3}{4}} \,\, \textrm{vol}_{5} \ ,
\end{equation}
where
\begin{equation}
\textrm{vol}_{5} = Y^{-\frac{3}{4}} \,\, \hat{e}^{5}\wedge \hat{e}^{6}\wedge \hat{e}^{7}\wedge \hat{e}^{8}\wedge \hat{e}^{9} \ ,
\end{equation}
is the volume form on the \textit{deformed} $\,\textrm{S}^5\,$ in (\ref{ds_6}). Finally the gauge-invariant five-form flux is given by
\begin{equation}
\label{Ftilde5_background}
\widetilde{F}_{5} =  dC + \tfrac{1}{2} \, \epsilon_{\alpha \beta} \, \mathbb{B}^{\alpha} \wedge \mathbb{H}^{\beta}  =  \left(4 +  \dfrac{6 \, (1-Y)}{Y} \right) \, Y^{\frac{3}{4}} \, (1 + \star) \, \textrm{vol}_{5} \ ,
\end{equation}
which breaks the $\,\textrm{U}(1)_{\mathbb{U}}\,$ symmetry whenever $\,Y \neq 1\,$. When particularised to the AdS$_{4}$ solutions obtained in the previous section the result is:
\begin{itemize}

\item[$i)\,$] For the $\,\mathcal{N}=1 \, / \, \textrm{SU}(3)\,$ solution in (\ref{vacuum_N=1_SK})-(\ref{vacuum_N=1_QK}) one has
\begin{equation}
\label{F5_potential_N1}
\begin{array}{lll}
\widetilde{F}_{5}  =  3 \, \left( \dfrac{6}{5}\right)^{\frac{3}{4}} \, (1 + \star) \, \textrm{vol}_{5} \ .
\end{array}
\end{equation}

\item[$ii)\,$] For the $\,\mathcal{N}=0 \, / \, \textrm{SO}(6)\,$ solution in (\ref{vacuum_N=0_SK})-(\ref{vacuum_N=0_QK}) one has
\begin{equation}
\label{F5_potential_N0}
\begin{array}{lll}
\widetilde{F}_{5}  =  4 \, (1 + \star) \, \textrm{vol}_{5} \ .
\end{array}
\end{equation}

\end{itemize}

\subsubsection*{Axion-dilaton and Janus}

The $\,\textrm{SL}(2)$-valued axion-dilaton $\,m_{\alpha \beta}\,$ of type IIB supergravity can be obtained from the last uplift formula in (\ref{uplift_formulas}). A straightforward computation involving this time the blocks of the $\,M^{MN}\,$ scalar matrix
\begin{equation}
\label{Mblocks_axion-dilaton}
\begin{array}{lll}
M^{\gamma\underline{k} \, \delta \underline{l}} &=&  e^{-\varphi} \, Y \,  \delta^{\underline{kl}} \,\,  \mathfrak{m}^{\gamma \delta} \, + \, e^{-\varphi} \, (1-Y) \, J^{\underline{kl}} \, \epsilon^{\gamma \delta} \ , \\[2mm]
M_{\gamma\underline{k} \, \delta \underline{l}}  &=&   e^{\varphi} \, Y \, \delta_{\underline{kl}} \,\, \mathfrak{m}_{\gamma \delta} \, + \, e^{\varphi} \, (1-Y) \, J_{\underline{kl}} \, \epsilon_{\gamma \delta}  \  ,
\end{array}
\end{equation}
yields an axion-dilaton of the form
\begin{equation}
\label{mab_matrix}
m_{\alpha \beta}= 
\frac{1}{\textrm{Im}\tau} \, \left(
\begin{array}{cc}
|\tau|^2 & - \textrm{Re}\tau  \\[2mm]
- \textrm{Re}\tau  &  1
\end{array}
\right)  
= (A^{-t})_{\alpha}{}^{\gamma} \, \mathfrak{m}_{\gamma \delta} \, (A^{-1})^{\delta}{}_{\beta} \ ,
\end{equation}
with $\,\tau=C_{0}+i \, e^{-\Phi}\,$. Note that the full dependence on the coordinate $\,\eta\,$ is again encoded into the matrix $\,A^{-1}(\eta)\,$ in (\ref{Ainv_twist}) which acts as an $\,\textrm{SO}(1,1) \subset \textrm{SL}(2)_{\textrm{IIB}}\,$ twist on 
\begin{equation}
\label{mathcal_m_matrix}
\mathfrak{m}_{\gamma \delta} \,\, = \,\, \frac{1}{Y}  \,  \, \left(
\begin{array}{ll}
e^{-2 \phi} \, (Y^2 + Z^2) & - Z \\[2mm]
-Z  & e^{2 \phi}
\end{array}
\right) \ .
\end{equation}

The matrix $\,\mathfrak{m}_{\gamma \delta}\,$ in (\ref{mathcal_m_matrix}) only depends on the four-dimensional scalars in the universal hypermultiplet. At both the $\,\mathcal{N}=1 \, / \, \textrm{SU}(3)\,$ and $\,\mathcal{N}=0 \, / \, \textrm{SO}(6)\,$ solutions one has that it reduces to
\begin{equation}
\label{mathcal_m_matrix_solutions}
\mathfrak{m}_{{\gamma \delta}} =  \frac{1}{\sqrt{1-\sigma^2}} \,  \, \left(
\begin{array}{ll}
1 & - \sigma \\[2mm]
-\sigma  & 1
\end{array}
\right)
\hspace{8mm} \textrm{ with } \hspace{8mm}
\sigma \in (-1,1) \ ,
\end{equation}
or, using an alternative parameterisation in terms of $\,\sigma = \tanh \mu\,$, to the manifest $\,\textrm{SO}(1,1)\,$ expression
\begin{equation}
\label{mathcal_m_matrix_solutions_mu}
\mathfrak{m}_{{\gamma \delta}} = \left(
\begin{array}{rr}
\cosh \mu & - \sinh \mu  \\[2mm]
 - \sinh \mu   & \cosh \mu
\end{array}
\right)
\hspace{8mm} \textrm{ with } \hspace{8mm}
\mu \in (-\infty,\infty) \ .
\end{equation}
Using (\ref{Ainv_twist}) and (\ref{mathcal_m_matrix_solutions_mu}) the axion-dilaton matrix in (\ref{mab_matrix}) takes the form
%
%
\begin{equation}
\label{m_matrix_solutions}
m_{\alpha \beta} = \left(
\begin{array}{rr}
\cosh (2 \, \eta + \mu ) & - \sinh (2 \, \eta + \mu )  \\[2mm]
 - \sinh (2 \, \eta + \mu )   & \cosh (2 \, \eta + \mu )
\end{array}
\right) \ ,
\end{equation}
and a direct comparison between (\ref{mab_matrix}) and (\ref{m_matrix_solutions}) allows us to extract the profile for the complex axion-dilaton field
%
%
\begin{equation}
\label{tau_general}
\textrm{Re} \tau = \tanh (2 \, \eta + \mu)
\hspace{8mm} , \hspace{8mm}
\textrm{Im} \tau =  \textrm{sech} (2 \, \eta + \mu) \ .
\end{equation}

In order to establish a connection with type IIB solutions of Janus-type \cite{Bak:2003jk,DHoker:2006vfr}, it will be convenient to combine the $\,A^{-1}(\eta)\,$ twist in (\ref{Ainv_twist}) with a (global) duality transformation $\,\Lambda \in \textrm{SO}(2) \subset \textrm{SL}(2)_{\textrm{IIB}}\,$ of the form
\begin{equation}
\label{Lambda_transf}
\Lambda_{\alpha}{}^{\gamma} = \frac{1}{\sqrt{2}} \, \left(
\begin{array}{cc}
 1 & -1 \\[2mm]
 1 & 1
\end{array}
\right) \ .
\end{equation}
Applying the transformation (\ref{Lambda_transf}) on our type IIB solutions, new but physically equivalent backgrounds are generated with
\begin{equation}
\label{m_new_solution_invariant}
\begin{array}{rll}
{ds^2_{10}}^{\textrm{(new)}} & = & ds^2_{10}\ , \\[3mm]
\widetilde{F}_{5}^{\,\textrm{(new)}} & = & \widetilde{F}_{5} \ , \\[3mm]
\mathbb{H}^{\alpha \, \textrm{(new)}} & = & (\mathbb{H} \, \Lambda^{-1} )^{\alpha} \ , 
\end{array}
\end{equation}
and
%
%
\begin{equation}
\label{m_new_solution}
\begin{array}{rll}
m_{\alpha \beta}^{\textrm{(new)}} &=& (\Lambda \, m \, \Lambda^{t})_{\alpha \beta} \,\,\,=\,\,\, 
\left(
\begin{array}{cc}
 e^{2 \eta + \mu} & 0 \\[2mm]
 0 & e^{- (2 \, \eta + \mu) }
\end{array}
\right) \ .
\end{array}
\end{equation}
In this way the composed action of $\,\Lambda \, A^{-1}(\eta)\,$ on (\ref{mathcal_m_matrix_solutions_mu}) reduces to just a shift of the form $\,\Phi^{\textrm{(new)}} \rightarrow \Phi^{\textrm{(new)}} -2 \eta \,$ and a Janus-like behaviour becomes manifest
\begin{equation}
e^{\Phi^{\textrm{(new)}}}=g^{\textrm{(new)}}_{s} = e^{- (2 \, \eta + \mu) }
\hspace{5mm} \Rightarrow \hspace{5mm}
\Phi^{\textrm{(new)}}(\eta) = - 2 \, \eta  - \mu \ ,
\end{equation}
with a linear dilaton $\,\Phi^{\textrm{(new)}}\,$ running from $\,-\infty\,$ to $\,\infty\,$ or, equivalently, taking two different asymptotic values one of them yielding a divergent string coupling $\,g^{\textrm{(new)}}_{s}\,$. Note that the free parameter $\,\sigma = \tanh \mu\,$ in the ten-dimensional solution, which appears as a modulus in four dimensions, can be used to set the dilaton to zero at an arbitrary value of~$\,\eta\,$. A natural choice is to set $\,\mu=0\,$ so that $\,\Phi^{\textrm{(new)}}(0)=0\,$. Lastly, we have also verified that the equations of motion and Bianchi identities of type IIB supergravity (see Appendix~\ref{App:Type_IIB}) are satisfied at both ten-dimensional $\,\textrm{SU}(3)\,$ and $\,\textrm{SO}(6)\,$ symmetric solutions.

\subsection{Summary of type IIB backgrounds}

By uplifting two families of AdS$_{4}$ vacua of the dyonically-gauged $\,{[\,\textrm{SO}(1,1) \times \textrm{SO}(6)\,] \ltimes \mathbb{R}^{12}}\,$ maximal supergravity, we have obtained two classes of ten-dimensional type IIB backgrounds.  In both classes the metric is non-singular and of the form $\,\textrm{AdS}_{4} \times \mathbb{R} \times \textrm{M}_{5}\,$ with $\,\eta \in (-\infty \, , \, \infty)\,$ being the coordinate along the $\,\mathbb{R}\,$ direction. The dependence of the backgrounds on the coordinate $\,\eta\,$ is fully encoded in an $\,\textrm{SO}(1,1) \subset \textrm{SL}(2)_{\textrm{IIB}}\,$ matrix
\begin{equation}
\label{Ainv_twist_summary}
A^{\alpha}{}_{\beta} =
\left(
\begin{array}{cc}
\cosh\eta & \sinh\eta \\[2mm]
\sinh\eta & \cosh\eta
\end{array}
\right)
\hspace{8mm} , \hspace{8mm}
(A^{-1})^{\alpha}{}_{\beta} =
\left(
\begin{array}{cc}
\cosh\eta & -\sinh\eta \\[2mm]
- \sinh\eta & \cosh\eta
\end{array}
\right) \ ,
\end{equation}
which acts as a twist on a \textit{constant} type IIB axion-dilaton
\begin{equation}
\label{mathcal_m_matrix_solutions_summary}
\mathfrak{m}_{{\gamma \delta}} =  \frac{1}{\sqrt{1-\sigma^2}} \,  \, \left(
\begin{array}{ll}
1 & - \sigma \\[2mm]
-\sigma  & 1
\end{array}
\right)
\hspace{8mm} \textrm{ with } \hspace{8mm}
\sigma \in (-1,1) \ ,
\end{equation}
and an $\textrm{SL}(2)_{\textrm{IIB}}$ doublet of $\eta$-independent three-form fluxes. From an effective $\mathcal{N}=2$ four-dimensional perspective, the free parameter $\,\sigma\,$ in the type IIB solutions corresponds to a four-dimensional axion in the universal hypermultiplet (see Section~\ref{sec:N=2_model}).
\\[-2mm]

\noindent We find two classes of type IIB backgrounds:

\begin{itemize}

\item[$i)$] The first class of solutions is $\,\mathcal{N}=1\,$ supersymmetric, and thus perturbatively stable, and preserves an $\,\textrm{SU}(3)\,$ symmetry arising from a $\,\mathbb{CP}^2 \subset \textrm{M}_{5}\,$ factor in the geometry. The various ten-dimensional fields are given by
\begin{equation}
\label{solution_N1_summary}
\begin{array}{lll}
ds_{10}^2 &=& \dfrac{3 \sqrt{6}}{10}\, ds^2_{\textrm{AdS}_{4}} + \dfrac{1}{3} \, \sqrt{\tfrac{10}{3}}  \, d\eta^{2} +  \left[   \, \sqrt{\dfrac{5}{6}} \, ds_{\mathbb{CP}^2}^2  +  \sqrt{\dfrac{6}{5}}  \, \boldsymbol{\eta}^2 \, \right] \ , \\[4mm]
\widetilde{F}_{5}  &=&  3 \left( \dfrac{6}{5}\right)^{\frac{3}{4}} \, (1 + \star) \, \textrm{vol}_{5} \ , \\[6mm]
m_{\alpha \beta} &=&  (A^{-t})_{\alpha}{}^{\gamma} \, \mathfrak{m}_{\gamma \delta} \, (A^{-1})^{\delta}{}_{\beta} \ , \\[4mm]
\mathbb{H}^{\alpha} &=&    A^{\alpha}{}_{\beta}  \,\, \mathfrak{h}^{\beta} \ ,
\end{array}
\end{equation}
with
\begin{equation}
\mathfrak{h}^{\beta} = -  \dfrac{5}{12}  \,  \epsilon^{\beta\gamma}  \, \left[ \, 3 \, H_{\gamma \delta} \, (i \, \boldsymbol{\Omega} \wedge \boldsymbol{\eta} )^{\delta} - \theta_{\gamma}{}^{\lambda} \,  H_{\lambda \delta} \, d\eta \wedge \boldsymbol{\Omega}^{\delta}\,  \right]  \ ,
\end{equation}
and where, in order to present $\,\mathbb{H}^{\alpha}\,$ in a concise form, we have introduced the two constant matrices
\begin{equation}
\theta_{\gamma}{}^{\lambda} =
\begin{pmatrix}
0 & 1  \\[2mm]
1  & 0 
\end{pmatrix} \ ,
\end{equation}
and
\begin{equation}
\label{Hmat_SU3_summary}
H_{\alpha \beta} = \frac{2 \sqrt{6}}{5 \, (1-\sigma ^2)^{\frac{1}{4}}}
\begin{pmatrix}
\sqrt{1-\sigma ^2}  \,\,  \cos\gamma  - \sigma \,\, \sin\gamma & \sqrt{1-\sigma ^2} \,\, \sin\gamma +  \sigma \,\, \cos\gamma  \\[2mm]
 \sin\gamma  & -\cos\gamma 
\end{pmatrix}  \ .
\end{equation}
The latter depends on the free parameter $\,\sigma \in (-1 , 1)\,$ specifying the constant axion-dilaton in (\ref{mathcal_m_matrix_solutions_summary}), as well as on an arbitrary angle $\,\gamma \in [0 , 2 \pi]\,$. As a result there is a $\,(\sigma,\gamma)$-family of three-form fluxes $\,\mathbb{H}^{\alpha}\,$. Note also that the internal geometry in (\ref{solution_N1_summary}) has an additional $\,\textrm{U}(1)_{\beta}\,$ isometry that is broken in the background by the dependence of the three-form fluxes $\,\mathbb{H}^{\alpha}\,$ on the complex $(2,0)$-form $\,\boldsymbol{\Omega}\,$ (see Appendix~\ref{App:Sasaki-Einstein}).

\item The second class of solutions is non-supersymmetric and preserves  an $\,\textrm{SO}(6)\,$ symmetry arising from an $\,\textrm{M}_{5}=\textrm{S}^{5}\,$ factor in the geometry. The various ten-dimensional fields are given by
\begin{equation}
\label{solution_N0_summary}
\begin{array}{lll}
ds_{10}^2 &=& \dfrac{1}{\sqrt{2}} \, ds^2_{\textrm{AdS}_{4}} + \dfrac{1}{2}  \, d\eta^{2} +  ds_{\textrm{S}^5}^2 \ , \\[6mm]
\widetilde{F}_{5}  &=&  4 \, (1 + \star) \, \textrm{vol}_{5} \ , \\[4mm]
m_{\alpha \beta} &=&  (A^{-t})_{\alpha}{}^{\gamma} \, \mathfrak{m}_{\gamma \delta} \, (A^{-1})^{\delta}{}_{\beta} \ , \\[3mm] 
\mathbb{H}^{\alpha} &=& 0 \ .
\end{array}
\end{equation}
This class of solutions features perturbative instabilities, as already noticed in (\ref{mass_N=0}) when looking at scalar fluctuations in the consistent truncation to a four-dimensional effective theory.

\end{itemize}

\subsection{S-fold interpretation}

Following the original reasoning in \cite{Inverso:2016eet} we argue now that the two AdS$_{4}$ solutions presented in this work actually uplift to S-fold backgrounds of type IIB supergravity. Since the matrix $\,\,m_{\alpha \beta}(\eta)\,$ in (\ref{m_matrix_solutions}) is of hyperbolic type, it is \textit{not} possible to perform a coordinate shift of the form $\,\eta \rightarrow \eta + T\,$ such that $\,m_{\alpha\beta}(\eta+T)=m_{\alpha\beta}(\eta)\,$. As a result the type IIB background cannot be viewed as a globally geometric compactification on $\,\textrm{S}^{1} \times \textrm{S}^{5}\,$.

However the $\,\eta\,$ direction on the internal geometry (\ref{ds_6}) can still be made periodic with period $\,T \neq 0\,$ at the price of introducing an $\,\textrm{SO}(1,1) \subset \textrm{SL}(2)_{\textrm{IIB}}\,$ hyperbolic monodromy $\,\mathfrak{M}_{\textrm{S}^{1}}\,$ on the axion-dilaton given by
\begin{equation}
\label{monodromy_S1}
\mathfrak{M}_{\textrm{S}^{1}} = 
A^{-1}(\eta) \, A(\eta+T) 
=
\left(
\begin{array}{rr}
 \cosh T & \sinh T \\
 \sinh T & \cosh T \\
\end{array}
\right) \ .
\end{equation}
The monodromy in (\ref{monodromy_S1}) can be brought to a generic hyperbolic (discrete) monodromy $\,\mathfrak{M}(k) \in  \textrm{SL}(2,\mathbb{Z})_{\textrm{IIB}}\,$ with $\,k\in\mathbb{Z}\,$ of the form
\begin{equation}
\label{monodromy_k}
\mathfrak{M}(k)=
\left( 
\begin{array}{ll}
k  & 1 \\[2mm]
-1  & 0
\end{array}
\right)
\hspace{8mm} \textrm{ , }\hspace{8mm}
k \ge 3 \ ,
\end{equation}
by performing a transformation
\begin{equation}
\mathfrak{M}(k)=g^{-1} \, \mathfrak{M}_{\textrm{S}^{1}} \, g
\hspace{10mm} \textrm{ with } \hspace{10mm}
g(k) = 
\left(
\begin{array}{cc}
 \dfrac{(k^2-4)^{\frac{1}{4}}}{\sqrt{2}} & 0 \\[2mm]
 \frac{k}{\sqrt{2} \, (k^2-4)^{\frac{1}{4}}} & \dfrac{\sqrt{2}}{(k^2-4)^{\frac{1}{4}}}
\end{array}
\right) \ .
\end{equation}
In this manner one has that
\begin{equation}
\mathfrak{M}(k) =  A^{-1}_{(k)}(\eta)  \,\,\, A_{(k)}\big(\eta+T(k)\big) \ ,
\end{equation}
with
\begin{equation}
A_{(k)}=A \, g(k)
\hspace{10mm} \textrm{ and } \hspace{10mm} 
T(k)=\textrm{ln} (k+\sqrt{k^2-4}) - \textrm{ln}(2) \ .
\end{equation}
The resulting type IIB background can in this way be interpreted as a locally geometric compactification on $\,\textrm{S}^{1} \times \textrm{S}^{5}\,$ with an S-duality monodromy $\,\mathfrak{M}(k)\in \textrm{SL}(2,\mathbb{Z})_{\textrm{IIB}}\,$ given by (\ref{monodromy_k}), namely, an S-fold background. To generate such a background one must replace $\, A^{-1}(\eta)  \rightarrow  A^{-1}_{(k)}(\eta) \,$ in the generalised Scherk--Schwarz anstaz, equivalently, in (\ref{(B,C)_potentials})-(\ref{H_background}) and (\ref{mab_matrix}). This procedure must also be composed with the global transformation in (\ref{Lambda_transf}) in order to make the Janus structure of the solution manifest. Moreover, since a generalised Scherk--Schwarz reduction satisfying the section constraint does not break supersymmetries and the monodromy in (\ref{monodromy_S1}) is of hyperbolic type, the supersymmetries preserved at the AdS$_{4}$ solutions uplift to supersymmetries of the ten-dimensional backgrounds\footnote{As discussed in \cite{Inverso:2016eet}, the fact that the monodromy in (\ref{monodromy_S1}) is of hyperbolic type and can be interpreted as a dilaton shift upon applying the (global) duality transformation in (\ref{Lambda_transf}), implies the existence of a global parameterisation of the axion-dilaton coset representative in (\ref{SL(2)_coset}) such that no (local) compensating $\,\textrm{SO}(2)\,$ transformation is induced by the monodromy on the type IIB fermions. Therefore, a Killing spinor at a supersymmetric AdS$_4$ solution uplifts to a globally well-defined Killing spinor in ten dimensions.}.

Finally it is to be noted that the class of monodromies in (\ref{monodromy_k}) can be expressed as a composition of $\,\mathcal{S}\,$ and $\,\mathcal{T}\,$ transformations. More concretely one has that
\begin{equation}
\mathfrak{M}(k) = -\mathcal{S} \, \mathcal{T}^{k} \ ,
\end{equation}
with
\begin{equation}
\mathcal{S}=\left( 
\begin{array}{cc}
0  & -1 \\[2mm]
1  & 0
\end{array}
\right)
\hspace{8mm} \textrm{ and }\hspace{8mm}
\mathcal{T}=\left( 
\begin{array}{cc}
1  & 0 \\[2mm]
1  & 1
\end{array}
\right) \ .
\end{equation}
This type of monodromies has been extensively studied in the context of three-dimensional $\,\mathcal{N}=4\,$ CFT known as $\,T(U(N))\,$ theories \cite{Gaiotto:2008sd,Gaiotto:2008ak}. It would be interesting to investigate general S-duality quotients of Janus-type non-compact solutions dual to three-dimensional $\,\mathcal{N}=1\,$ S-fold CFTs (see \cite{Assel:2018vtq} for a study of $\,\mathcal{N}=4\,$ S-fold CFTs) as well as their potential realisation in terms of brane configurations.

\section{Conclusions}
\label{sec:conclusions}

Fetching techniques from $\,\textrm{E}_{7(7)}$-EFT and generalised Scherk--Schwarz reductions, in this work we have uplifted two  different AdS$_{4}$ solutions of the four-dimensional and dyonically-gauged maximal supergravity with $\,{[\,\textrm{SO}(1,1) \times \textrm{SO}(6)\,] \ltimes \mathbb{R}^{12}}\,$ gauge group to two classes of S-fold backgrounds of type IIB supergravity. These S-folds result from $\,\textrm{SL}(2,\mathbb{Z})_{\textrm{IIB}}\,$ quotients of Janus-type non-compact solutions of the form $\,\textrm{AdS}_{4} \times \mathbb{R} \times \textrm{M}_{5}\,$. The first class of S-folds preserves $\,\mathcal{N}=1\,$ supersymmetry and an $\,\textrm{SU}(3)\,$ symmetry originating from an internal geometry of the form $\,\textrm{M}_5 = \mathbb{CP}_{2} \rtimes \textrm{S}^1\,$. The additional $\,\textrm{U}(1)_{\beta}\,$ symmetry associated with $\,\textrm{S}^1\,$ is broken due to the presence of three-form fluxes. The second class of S-folds is non-supersymmetric and features an $\,\textrm{SO}(6)\,$ symmetry as a consequence of the internal geometry $\,\textrm{M}_{5}=\textrm{S}^{5}\,$ being a round five-sphere.

Both classes of S-folds are obtained from Janus-like solutions with a linear dilaton profile upon quotients by hyperbolic elements of $\,\textrm{SL}(2,\mathbb{Z})_{\textrm{IIB}}\,$. This translates into a monodromy on the axion-dilaton of the form $\,-\mathcal{S} \, \mathcal{T}^{k}\,$ with $\,k \ge 3\,$. While supersymmetry guarantees the stability of the $\,\mathcal{N}=1\,$ S-folds with $\,\textrm{SU}(3)\,$ symmetry, the issue of stability becomes more subtle in the absence of supersymmetry. The non-supersymmetric Janus-like non-compact solution with a linear dilaton and $\,\textrm{SO}(6)\,$ symmetry presented in this work is of the ``curious" type discussed in \cite{Freedman:2003ax} (see also \cite{Robb:1984uj}). There the issue of stability was left open for this particular type of solutions. Here, by performing an analysis of normalised scalar masses in the effective four-dimensional gauged supergravity, we observe the presence of unstable modes in (\ref{mass_N=0}) violating the BF bound for perturbative stability in AdS$_{4}\,$. The consistency of the truncation of type IIB supergravity on $\,\mathbb{R} \times \textrm{S}^{5}\,$ down to a maximal $D=4$ supergravity then renders the non-supersymmetric type IIB background also unstable.

The $\,\mathcal{N}=1\,$ S-folds with $\,\textrm{SU}(3)\,$ symmetry presented in this work pair up with the ten-dimensional massive IIA solution featuring the same (super) symmetries presented in \cite{Varela:2015uca}. It would be interesting to further investigate a possible connection between the two classes of type II solutions. On the contrary, the higher-dimensional origin (if any) of the analogous AdS$_{4}$ solution in the dyonically-gauged $\,\textrm{SO}(8)\,$ theory (see Table~\ref{Table:vacua}) remains elusive, and a no-go theorem has been proved against its existence as a compactification that is locally described by eleven-dimensional supergravity \cite{Lee:2015xga}. This makes the (four-dimensional) phenomenon of electromagnetic duality in maximal supergravity more mysterious from an \mbox{M-theory} perspective.

As a final point, it would be interesting to investigate possible brane setups underlying the $\,\mathcal{N}=1\,$ S-folds with $\,\textrm{SU}(3)\,$ symmetry, and to explore the associated dual three-dimensional $\,\mathcal{N}=1\,$ S-fold CFTs. Since an $\,\mathcal{N}=1 \, / \, \textrm{SU}(3)\,$ AdS$_{4}$ solution occurs in the three dyonically-gauged maximal supergravities of Table~\ref{Table:vacua} with exactly the same normalised scalar mass spectrum, a better understanding of the holographic aspects of the type IIA/IIB solutions could help in getting new insights into the dyonically-gauged $\,\textrm{SO}(8)\,$ theory from a novel holographic perspective. We hope to come back to these and other related issues in the future.

\section*{Acknowledgements}

We would like to thank Riccardo Argurio, Andr\'es Collinucci, Carlos Hoyos and Patrick Meessen for conversations and, especially, Mario Trigiante for initial collaboration on this project and helpful comments. The work of AG is partially supported by the Spanish government grant MINECO-16-FPA2015-63667-P, by the Principado de Asturias through the grant FC-GRUPIN-IDI/2018/000174, by the ERC Advanced Grant ``High-Spin-Grav" and by F.R.S.-FNRS through the conventions PDRT.1025.14 and IISN-4.4503.15. AG would like to acknowledge the Universit\'e Libre de Bruxelles where this work was initiated as well as the Mainz Institute for Theoretical Physics (MITP) of the Cluster of Excellence PRISMA+ (Project ID 39083149) where this work was completed.

\appendix

\section{Canonical Sasaki-Einstein structure on $\,\textrm{S}^5\,$}
\label{App:Sasaki-Einstein}

The connection between the various components of the scalar matrix $\,M^{MN}\,$ and the $\,\textrm{SU}(2)$-structure of the five-sphere $\,\textrm{S}^5\,$ when viewed as a Sasaki-Einstein manifold relies on the $\,\mathbb{R}^6\,$ embedding relations
\begin{equation}
\label{SU2_structure}
\boldsymbol{\eta} = J_{\underline{mn}} \, \mathcal{Y}^{\underline{m}} \, d\mathcal{Y}^{\underline{n}}
\hspace{5mm} , \hspace{5mm}
\boldsymbol{J} = \frac{1}{2} \, J_{\underline{mn}} \, d\mathcal{Y}^{\underline{m}} \wedge d\mathcal{Y}^{\underline{n}}
\hspace{5mm} , \hspace{5mm}
\boldsymbol{\Omega} = \frac{1}{2} \, \Omega_{\underline{mnp}} \, \mathcal{Y}^{\underline{m}} \, d\mathcal{Y}^{\underline{n}} \wedge d\mathcal{Y}^{\underline{p}} \ ,
\end{equation}
determining the real one-form $\,\boldsymbol{\eta}\,$, real two-form $\,\boldsymbol{J}\,$ and complex $(2,0)$-form $\,\boldsymbol{\Omega}\,$ of the $\,\textrm{SU}(2)$-structure in terms of the real two-form $\,J\,$ and the holomorphic three-form $\,\Omega\,$ of the $\textrm{SU}(3)$-holonomy of the K\"ahler cone $\,C(\textrm{S}^5)=\mathbb{R}^{+} \times \textrm{S}^5\,$. The latter are given by
\begin{equation}
\begin{array}{rlll}
J &=& e^{2} \wedge e^{3} + e^{4} \wedge e^{5} + e^{6} \wedge e^{7} & , \\[2mm]
\Omega &=& \Omega^{R} + i \,  \Omega^{I} = (e^{2}+ i \, e^{3}) \wedge (e^{4}+ i \, e^{5}) \wedge  (e^{6}+ i \, e^{7}) & .
\end{array}
\end{equation}
The closure of $\,J\,$ and $\Omega$ translates into the torsion conditions for the $\,\textrm{SU}(2)$-structure
\begin{equation}
\label{dSU(2)-structure}
d\boldsymbol{\eta} = \boldsymbol{J}
\hspace{5mm} , \hspace{5mm}
d\boldsymbol{J} = 0
\hspace{5mm} , \hspace{5mm}
d\boldsymbol{\Omega} = 3 \, i \, \boldsymbol{\eta}  \wedge \boldsymbol{\Omega} \ .
\end{equation}

The metric on the round $\,\textrm{S}^5\,$ can be expressed (locally) as a $\,\textrm{U}(1)_{\beta}\,$ fibration with coordinate $\,\beta\,$ over the K\"ahler-Einstein space $\,\mathbb{CP}^2\,$ with metric
\begin{equation}
\label{metric_CP2xU1}
ds_{\textrm{S}^5}^2 = ds_{\mathbb{CP}^2}^2 + \boldsymbol{\eta}^2 \ ,
\end{equation}
where $\,\boldsymbol{\eta}=d\beta + \boldsymbol{A}_{1}\,$ and $\,\boldsymbol{A}_{1}\,$ is the one-form potential on $\,\mathbb{CP}^2\,$ such that $\,d\boldsymbol{A}_{1}=2 \, \boldsymbol{J}\,$. Following the conventions in \cite{DHoker:2006vfr} we (locally) choose angular coordinates on $\,\textrm{S}^5 \sim \mathbb{CP}_{2} \rtimes \textrm{S}^1\,$ such that the coordinates $\,y^{i}\,$ take the parametric form
\begin{equation}
\begin{array}{lll}
y^{2} &=& \sin\alpha \, \cos\frac{\theta}{2} \, \cos(\frac{\psi+\phi}{2}+\beta) \ , \\[2mm]
y^{3} &=& \sin\alpha \, \cos\frac{\theta}{2} \, \sin(\frac{\psi+\phi}{2}+\beta) \ , \\[2mm]
y^{4} &=& \sin\alpha \, \sin\frac{\theta}{2} \, \cos(\frac{\psi-\phi}{2}+\beta) \ , \\[2mm]
y^{5} &=& \sin\alpha \, \sin\frac{\theta}{2} \, \sin(\frac{\psi-\phi}{2}+\beta) \ , \\[2mm]
y^{6} &=& - \cos\alpha \, \sin\beta \ ,
\end{array}
\end{equation}
so that 
\begin{equation}
\hat{\rho}^2 = \sqrt{1-|\vec{y}|^2} = \cos\alpha \,  \cos\beta
\hspace{8mm} \textrm{ and } \hspace{8mm}
\mathring{\rho}^2 = \sqrt{1 + \tilde{y}^2} = \cosh\eta \ .
\end{equation}
The metric on $\,\mathbb{CP}^2\,$ takes the form
\begin{equation}
\label{metric_CP2}
ds^{2}_{\mathbb{CP}_{2}} = \delta_{ab} \, \hat{e}^{a}  \, \hat{e}^{b}
\hspace{8mm} \textrm{ with } \hspace{8mm}
a=5, 6, 7, 8 \ ,
\end{equation}
when expressed in terms of the frame fields
\begin{equation}
\hat{e}^{5} = d \alpha
\hspace{5mm} , \hspace{5mm}
\hat{e}^{6} = \tfrac{1}{4} \, \sin(2\alpha) \,  \sigma_{3}
\hspace{5mm} , \hspace{5mm}
\hat{e}^{7} = \tfrac{1}{2} \, \sin(\alpha) \,  \sigma_{1}  
\hspace{5mm} , \hspace{5mm}
\hat{e}^{8} = \tfrac{1}{2} \, \sin(\alpha) \,  \sigma_{2} 
\end{equation}
with
\begin{equation}
\begin{array}{rll}
\sigma_{1} &=& -\sin\psi \, d\theta + \cos \psi \, \sin \theta \, d \phi \ , \\[2mm]
\sigma_{2} &=& \cos \psi \, d\theta + \sin \psi \, \sin \theta \, d \phi \ , \\[2mm]
\sigma_{3} &=& d\psi + \cos \theta \, d \phi \ ,
\end{array}
\end{equation}
being a set of $\,\textrm{SU}(2)\,$ left-invariant forms. A direct substitution yields
\begin{equation}
\label{metric_CP2_1}
ds^{2}_{\mathbb{CP}_{2}} = d \alpha^2 + \frac{1}{4} \, \sin^2 \alpha \left( \sigma_{1}^2 + \sigma_{2}^2 +  \cos^2\alpha \, \sigma_{3}^2 \right) \ .
\end{equation}
Moreover the metric in (\ref{metric_CP2_1}) can be brought into a Fubini-Study form
\begin{equation}
\label{metric_CP2_FS}
ds^{2}_{\mathbb{CP}_{2}} = g_{i\bar{j}} \, dz^{i}  \, d\bar{z}^{\bar{j}}
\hspace{8mm} \textrm{ with } \hspace{8mm}
g_{i\bar{j}} = \partial_{i} \partial_{\bar{j}} \, \ln\left(1+ |z_{1}|^2+ |z_{2}|^2 \right) \ ,
\end{equation}
by introducing two complex coordinates
\begin{equation}
z_{1} = \tan \alpha \,  \cos\left( \frac{\theta}{2}\right) \, e^{i\frac{(\psi+\phi)}{2}}
\hspace{8mm} \textrm{ and } \hspace{8mm}
z_{2} = \tan \alpha \,  \sin\left( \frac{\theta}{2}\right) \, e^{i\frac{(\psi-\phi)}{2}} \ ,
\end{equation}
in terms of which the one-form potential $\,\boldsymbol{A}_{1}\,$ on $\,\mathbb{CP}^2\,$ reads
\begin{equation}
\boldsymbol{A}_{1} = - \frac{i}{2} \, \frac{\bar{z}_{i} \, d z_{i} - z_{i} \, d \bar{z}_{i}}{1+ |z_{1}|^2+ |z_{2}|^2} = \frac{1}{2} \, \sin^2 \alpha \, \sigma_{3} \ .
\end{equation}
As a result the set of $\,\textrm{SU}(3)\,$ invariant forms on $\,\textrm{S}^5\,$ is given by
\begin{equation}
\label{SU(3)-structure_S5}
\begin{array}{rlll}
\boldsymbol{\eta} &=& d\beta + \boldsymbol{A}_{1} \,\,=\,\, d\beta + \tfrac{1}{2} \, \sin^2 \alpha \, \sigma_{3} & , \\[2mm]
\boldsymbol{J} &=& d \boldsymbol{\eta} \,\,=\,\, 2 \, ( \hat{e}^{5} \wedge \hat{e}^{6} + \hat{e}^{7} \wedge \hat{e}^{8} ) & , \\[2mm]
\boldsymbol{\Omega} &=&  e^{3 i \beta} \,  (\hat{e}^{5} + i \, \hat{e}^{6}) \wedge (\hat{e}^{7} + i \, \hat{e}^{8}) & ,
\end{array}
\end{equation}
and satisfy the algebraic relations
\begin{equation}
\boldsymbol{\Omega} \wedge \bar{\boldsymbol{\Omega}} = \tfrac{1}{2} \, \boldsymbol{J} \wedge \, \boldsymbol{J} = 4 \, \hat{e}^{5} \wedge  \hat{e}^{6} \wedge  \hat{e}^{7} \wedge  \hat{e}^{8}
\hspace{8mm} \textrm{ and } \hspace{8mm}
\boldsymbol{J} \wedge \boldsymbol{\Omega} = 0 \ .
\end{equation}
Note that $\,\boldsymbol{\Omega}\,$ in (\ref{SU(3)-structure_S5}) transforms with a phase under $\,\textrm{U}(1)_{\beta}\,$ whereas $\,\boldsymbol{J}\,$ and $\,\boldsymbol{\eta}\,$ are neutral. Lastly, using (\ref{metric_CP2}), the metric (\ref{metric_CP2xU1}) on the round $\,\textrm{S}^{5}\,$ can be expressed as
\begin{equation}
\label{metric_S5}
ds^{2}_{\textrm{S}^5} = \delta_{ab} \, \hat{e}^{a} \, \, \hat{e}^{b} + \hat{e}^{9} \, \hat{e}^{9} 
\hspace{10mm} \textrm{ with } \hspace{10mm}
\hat{e}^{9}=\boldsymbol{\eta} \ .
\end{equation}

\section{Type IIB supergravity}
\label{App:Type_IIB}

The bosonic massless spectrum of ten-dimensional (chiral) type IIB supergravity  contains -- besides the universal NS-NS sector that includes the metric $\,G\,$, a two-form $\,B_2\,$ with field strength $\,H_{3}=dB_{2}\,$, and the dilaton $\,\Phi\,$ -- a set of even $p$-forms in the R-R sector. In particular, a fourth-rank antisymmetric self-dual tensor $\,C_{4}$, a two-form $\,C_{2}\,$ and a scalar $\,C_{0}$. The bosonic part of the type IIB supergravity action in the Einstein's frame consists of the three terms\footnote{Our conventions are related to those of \cite{Cremmer:1997ct} by the rescaling $\,C_{4}{}^{\cite{Cremmer:1997ct}} = \frac{1}{\sqrt{2}} \, C_{4}\,$. Moreover, we define $\,|H_{3}|^2 \equiv \frac{1}{3!} \, H_{MNP} \, H^{MNP}\,$ and similarly for the R-R field strengths.}
\begin{equation}
\label{S_bos2}
S_{bos} = S_{\textrm{NS-NS}} + S_{\textrm{R-R}} + S_{\textrm{CS}} \ .
\end{equation}
It contains the $\,S_{\textrm{NS-NS}}\,$ term accounting for the fields in the universal sector, namely, 
\begin{equation}
\label{action_NS-NS_IIB}
S_{\textrm{NS-NS}}  =  \dfrac{1}{2 \kappa^{2}}  \int d^{10}x \, \sqrt{-G} \, \left(   R - \tfrac{1}{2} \, \partial^{M} \Phi \partial_{M} \Phi  - \tfrac{1}{2} e^{-\Phi}  |H_{3}|^{2}  \right) \ .
\end{equation}
The $\,S_{\textrm{R-R}}\,$ term in the action controlling the dynamics of the R-R fields $\,C_{0}$, $\,C_{2}\,$ and $\,C_{4}\,$ is given by
\begin{equation}
\label{action_R-R_IIB}
S_{\textrm{R-R}}  =  \dfrac{1}{2 \kappa^{2}}  \int d^{10}x \, \sqrt{-G}  \, \left(  -\tfrac{1}{2}\,  e^{2 \Phi}  |F_{1}|^{2}  -\tfrac{1}{2}\,  e^{\Phi}  |\widetilde{F}_{3}|^{2}  -\tfrac{1}{4}\, |\widetilde{F}_{5}|^{2} \right)  \ , 
\end{equation}
where the tilded field strengths are defined as
\begin{equation}
\label{widetildeFs}
\begin{array}{ccl}
\widetilde{F}_{3} & = & F_{3} - C_{0} \wedge H_{3} \ , \\
\widetilde{F}_{5} & = & F_{5} + \tfrac{1}{2} \, \left( B_{2} \wedge F_{3} - C_{2} \wedge H_{3}  \right) \ ,
\end{array}
\end{equation}
in terms of the standard ones $\,F_{n+1}=dC_{n}\,$. Additionally, the self-duality condition 
\begin{equation}
\widetilde{F}_{5} = \star \widetilde{F}_{5} \ ,
\end{equation}
with $\,(\star \widetilde{F})^{MNOPQM} \equiv \dfrac{1}{5!\,\sqrt{-G}} \, \epsilon^{MNOPQM'N'O'P'Q'} \, \widetilde{F}_{M'N'O'P'Q'} \,$ has to be supplemented by hand in order to have the correct number of bosonic degrees of freedom. 
The type IIB theory also incorporates a topological Chern-Simons term $\,S_{\textrm{CS}}\,$ in the action given by 
\begin{equation}
\label{action_CS_IIB}
S_{\textrm{CS}}  =  -\frac{1}{4 \,  \kappa^{2}}  \int   C_{4} \wedge  H_{3} \wedge F_{3}  \ .
\end{equation}

The equations of motion that follow from the action (\ref{S_bos2}) with the various contributions in (\ref{action_NS-NS_IIB}), (\ref{action_R-R_IIB}) and (\ref{action_CS_IIB}) are given by
\begin{equation}
\label{pforms_EOM}
\begin{array}{rll}
d \star \widetilde{F}_{5} & = & \frac{1}{2} \, \epsilon_{\alpha \beta} \, \widetilde{\mathbb{H}}^{\alpha} \wedge  \widetilde{\mathbb{H}}^{\beta} \ , \\[2mm]
d \star (e^{-\Phi} \, H_{3} - e^{\Phi} \, C_{0} \, \widetilde{F}_{3} ) & = & - \widetilde{F}_{5} \wedge (\widetilde{F}_{3} + C_{0} \, H_{3}) \ , \\[2mm] 
d \star (e^{\Phi} \, \widetilde{F}_{3} ) & = &  \widetilde{F}_{5} \wedge H_{3}\ ,  \\[2mm] 
\nabla^{M} (e^{2 \Phi} \, \partial_{M} C_{0} ) & = & - \frac{1}{3!} \, e^{\Phi} \, H_{MNP} \, \widetilde{F}^{MNP} \ , \\[2mm] 
\square\Phi & = & e^{2\Phi} \, |F_{1}|^2 + \frac{1}{2} \, e^{-\Phi} \, 	|H_{3}|^2 -\frac{1}{2} \, e^{\Phi} \, 	|\widetilde{F}_{3}|^2 \ ,
\end{array}
\end{equation}
where $\,\widetilde{\mathbb{H}}^{\alpha} = (H_{3},\widetilde{F}_{3})\,$ and $\,\square\Phi \equiv \nabla^{M} \partial_{M} \Phi \,$, together with the Einstein equation
\begin{equation}
\label{Einstein_EOM}
\begin{array}{rll}
R_{MN} &=& \frac{1}{2} \, \partial_{M} \Phi \,  \partial_{N} \Phi +
\frac{1}{2} \, e^{2\Phi}\, \partial_{M} C_{0} \,  \partial_{N} C_{0} \\[2mm]
&+& \frac{1}{4} \, \frac{1}{4!} \left(\widetilde{F}_{M P_{1}\cdots P_{4}}  \, \widetilde{F}_{N}{}^{P_{1}\cdots P_{4}} - \frac{1}{10} \, \widetilde{F}_{P_{1}\cdots P_{5}}  \, \widetilde{F}^{P_{1}\cdots P_{5}} \, G_{MN} 	\right) \\[2mm]
&+& \frac{1}{4} \, e^{-\Phi} \, \left( H_{M P_{1}P_{2}}  \, H_{N}{}^{P_{1}P_{2}} - \frac{1}{12} \, H_{P_{1}P_{2}P_{3}}  \, H^{P_{1}P_{2}P_{3}} \, G_{MN} 	\right) \\[2mm]
&+& \frac{1}{4} \, e^{\Phi} \, \left( \widetilde{F}_{M P_{1}P_{2}}  \, \widetilde{F}_{N}{}^{P_{1}P_{2}} - \frac{1}{12} \, \widetilde{F}_{P_{1}P_{2}P_{3}}  \, \widetilde{F}^{P_{1}P_{2}P_{3}} \, G_{MN} 	\right) \ .
\end{array}
\end{equation}
In addition, the set of Bianchi identities for the various gauge potentials reads
\begin{equation}
\label{pforms_BI}
dH_{3} = 0
\hspace{5mm} , \hspace{5mm}
dF_{1} = 0
\hspace{5mm} , \hspace{5mm}
d\widetilde{F}_{3} = - F_{1} \wedge H_{3}
\hspace{5mm} , \hspace{5mm}
d\widetilde{F}_{5} = H_{3} \wedge F_{3} \ .
\end{equation}
Note the equivalence between the first equation of motion in (\ref{pforms_EOM}) and the last Bianchi identity in (\ref{pforms_BI}) for the self-dual $\,C_{4}\,$ potential.

The action (\ref{S_bos2}) has a global $\,\textrm{SL}(2)_{\textrm{IIB}}\,$ invariance which becomes manifest when combining the axion $\,C_{0}\,$ and the dilaton $\,\Phi\,$ into a  coset representative $\,\mathcal{V}_{2} \in \textrm{SL}(2)_{\textrm{IIB}}/\textrm{SO}(2)\,$ such that the axion-dilaton matrix $\,m_{\alpha \beta}\,$ reads
\begin{equation}
\label{SL(2)_coset}
m_{\alpha \beta} = (\mathcal{V}_{2} \, {\mathcal{V}_{2}}^{t})_{\alpha \beta} = e^{\Phi} \, \left(
\begin{array}{cc}
e^{-2\Phi}  +  C_{0}{}^2 & - C_{0}  \\[2mm]
- C_{0}  &  1
\end{array}
\right)  \ .
\end{equation}
In terms of this matrix the second and third equations of motion in (\ref{pforms_EOM}) are re-expressed in an $\,\textrm{SL(2)}_{\textrm{IIB}}\,$ covariant form
\begin{equation}
d \star (m_{\alpha \beta}\, \mathbb{H}^{\beta}) = -  \epsilon_{\alpha\beta}  \, \widetilde{F}_{5} \wedge \mathbb{H}^{\beta} \ ,
\end{equation}
where $\,\mathbb{H}^{\alpha} = (H_{3},F_{3})\,$.

\bibliography{references}

\end{document}